\documentclass[12pt]{article}
\usepackage{latexsym}
\usepackage{amsmath}
\usepackage{amssymb}

\usepackage{verbatim}
\usepackage{bm}

\usepackage{curves}	
\usepackage{epic}
\usepackage{eepic}
\usepackage{epsfig}

\newcommand{\dd}{\mbox{\rm d}}
\newcommand{\wg}{\wedge}
\newcommand{\gam}{\gamma}

\newcommand{\dg}{\dagger}
\newcommand{\ddg}{\ddagger}
\newcommand{\tl}{\tilde}

\newcommand{\hmu}{{\hat \mu}}
\newcommand{\hnu}{{\hat \nu}}

\newcommand{\nnn}{\noindent}

\newcommand{\p}{\partial}

\newcommand{\be}{\begin{equation}}
\newcommand{\bear}{\begin{eqnarray}}
\newcommand{\ear}{\end{eqnarray}}
\newcommand{\ee}{\end{equation}}

\newcommand{\lbl}{\label}
\newcommand{\bi}{\bibitem}
\newcommand{\ci}{\cite}

\newcommand{\vs}{\vspace}
\newcommand{\hs}{\hspace}

\newcommand{\bD}{{\bar D}}
\newcommand{\vphi}{\varphi}
\newcommand{\sg}{\sigma}
\newcommand{\vac}{|0 \rangle}
\newcommand{\ba}{{\bar a}}
\newcommand{\bb}{{\bar b}}

\newcommand{\bmu}{{\bar \mu}}

\newcommand{\bp}{{\bm p}}
\newcommand{\bk}{{\bm k}}
\newcommand{\bx}{{\bm x}}
\newcommand{\hbp}{{\hat{\bm p}}}

\begin{document}

\begin{center}

\

\vs{.8cm}

\baselineskip .7cm

{\bf \Large Branes and Quantized Fields} 

\vs{4mm}

\baselineskip .5cm
Matej Pav\v si\v c

Jo\v zef Stefan Institute, Jamova 39,
1000 Ljubljana, Slovenia

e-mail: matej.pavsic@ijs.si

\vs{3mm}

{\bf Abstract}
\end{center}

\baselineskip .43cm
{\footnotesize
It is shown that the Dirac-Nambu-Goto brane can be described as a point particle in
an infinite dimensional space with a particular metric. This can be considered as a special case of a general theory in which branes are points in the brane space ${\cal M}$, whose metric is dynamical, just like in general relativity. Such a brane theory, amongst others, includes the flat brane space, whose metric is the infinite dimensional analog of the Minkowski space metric $\eta_{\mu \nu}$. A brane living in the latter space will be called ``flat brane''; it is like a bunch of non-interacting point particles. Quantization of the latter system leads to a system of non-interacting quantum fields. Interactions can be included if we consider a non trivial metric in the space of fields. Then the effective classical brane is no longer a flat brane. For a particular choice of the metric in the field space we obtain the Dirac-Nambu-Goto brane. We also show how a Stueckelberg-like quantum field arises within the brane space formalism. With the Stueckelberg fields, we avoid certain well-known intricacies, especially those related to the position operator that is needed in our construction of effective classical branes from the systems of quantum fields.}

\vs{3mm}

\hs{7mm}

\baselineskip .55cm

\section{Introduction}

Relativistic membranes of arbitrary dimension (branes)\,\ci{pBranes}--\ci{DuffBenchmark},
are very important objects in theoretical physics.
An attractive possibility is a brane world scenario\,\ci{BraneWorld}--\ci{PavsicBrane3}
in which  our spacetime is a 4-dimensional surface
embedded in a higher dimensions space. Quantization of gravity could then be achieved by
quantizing the brane. Unfortunately, quantization of the Dirac-Nambu-Goto brane, satisfying
the minimal surface action principle, is a tough problem that has not yet been solved in general.
Although the quantization of the string, an extended object whose worldsheet has two dimensions,
is rather well understood\,\ci{strings}--\ci{strings3}, this is not so in the case  of branes
with higher dimensional worldsheets (also called ``worldvolumes").

We will show how to solve this problem by considering the brane as a point in an infinite-dimensional
brane space ${\cal M}$ that in general can be curved. The idea is that the metric of ${\cal M}$ is dynamical,
just like in general relativity\,\ci{PavsicBook,PavsicPortoroz,PavsicNewBrane}.
In particular the metric of ${\cal M}$
can be such that it gives the {\it Dirac-Nambu-Goto brane}, which is just the usual ``minimal surface'' brane.
For other choices of ${\cal M}$-space metric we have branes that differ from the Dirac-Nambu-Goto brane,
i.e., they do not satisfy the minimal surface action principle, but some other action principle. In particular,
the ${\cal M}$-space metric can be ``flat", which means that at any point of ${\cal M}$ it can be cast
into the diagonal form. Then we have a brane analogue of a point particle in flat spacetime. Such a brane,
from now on called {\it flat brane}, sweeps a worldsheet that is a bunch of straight worldlines (Fig.\,1).
\setlength{\unitlength}{.8mm}

\begin{figure}[h!]
\hs{3mm} \begin{picture}(120,47 )(0,0)
\put(60,0){\includegraphics[scale=0.6]{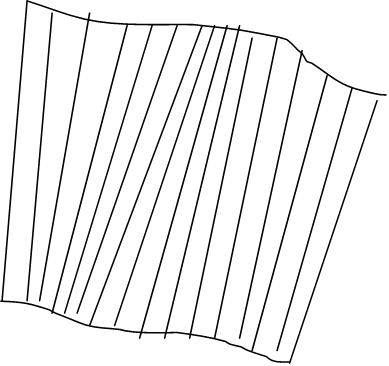}}
\put(75,0){$\sg$}
\put(58,27){$\tau$}

\end{picture}

\caption{\footnotesize  A schematic illustration of a ``flat brane".  For some exact plots see Fig.\,2.  }
\end{figure} 

A flat brane is thus just like continuous system of point particles. Quantization of a flat brane then leads
to a system of non interacting quantum fields, $\varphi^r (x)$. The index $r$ distinguishes one field from
another, and because in the classical theory we had a continuous set of point particles, $r$ must be
continuous. If we consider interactions among the quantum fields, $\lambda_{rs} \vphi^r (x) \vphi^s (x)$,
then the effective classical theory gives a brane living in a curved brane space, which,
in particular, can be the Dirac-Nambu-Goto brane\,\ci{PavsicNewBrane}.

Instead of one brane a system can consist of many branes.  Within the framework
of such an enlarged configuration space it is possible to formulate the Stueckelberg quantum field theory
with an invariant evolution parameter. We show how the latter parameter is embedded in the system's
configuration.

Each of those branes can be described
with a finite number of degrees of freedom, namely, with the center of mass, and additional
degrees of freedom that take into account finite extension of the brane. Such extra degrees
of freedom can be the coordinates of the Clifford space\,\ci{Castro1}--\ci{CastroPavsicRev} that includes scalars,
oriented lengths, areas,
volumes and 4- volumes (pseudoscalars). In describing a multi brane system
we can choose one brane and sample
it with the coordinates of Clifford space, while for the remaining branes we retain the description
with embedding functions. Clifford space is a 16-dimensional ultrahyperbolic space with
neutral signature (8,8). From the scalar and pseudoscalar coordinates that span a 2-dimensional
subspace with signature $(+ -)$, we can form, with a suitable superposition, the analog of
the light cone coordinates. In ultrahyperbolic spaces the Cauchy problem in general cannot
be well posed, unless we determine initial data on the ``light cone".  In such a way we obtain
the generalized Stueckelberg description\,\ci{StueckelbergField1}--\ci{StueckelbergField11},
\ci{PavsicBook}
of particles and branes, both classical and quantum.
The Stueckelberg theory is based on the introduction of an evolution parameter $\tau$, which
is invariant under Lorentz transformations. In the literature we can find various explanations
about the physical origin of $\tau$, but none is generally accepted. In the approach
pursued in this and in a series of previous papers\,\ci{PavsicBook,PavsicHualien}, the evolution parameter
$\tau$ is a superposition of the scalar and pseudoscalar coordinate of the Clifford space, and
is thus embedded in the configuration of the chosen brane. The latter brane, which in fact
need not be just a brane, but whatever extended object that can be sampled as a brane,
thus serves as a clock with which we measure the motion of the remaining branes that form
the considered system.

\section{Brane as a point in the brane space ${\cal M}$}

The Dirac-Nambu-Goto brane is described by the minimal surface action
\be
  I =   \kappa \int \dd^{p+1} \xi \, (-\gam )^{1/2},
\lbl{2.1}
\ee
where $\gam \equiv {\rm det} \,\gam_{ab}$, $\gam_{ab} \equiv \p_a X^\mu \p_b X_\mu$
is the determinant of the worldsheet embedding functions $X^\mu (\xi^a)$, $\mu=0,1,2,...,D-1$,
$a=0,1,2,...,p$.

An equivalent action is the Schild action\,\ci{Schild}
\be
  I_{\rm Schild} =\frac{\kappa}{2 k} \int \dd^{p+1}\xi \, (-\gam),
\lbl{2.2}
\ee
from which as a consequence of equations of motion we obtain
\be
\p_a (-\gam) =0.
\lbl{2.3}
\ee
The determinant of the induced metric $\gam$ is thus a constant whose choice determines
a gauge. We will choose a gauge such that
\be
-\gam = k^2,
\lbl{2.3a}
\ee
 in which case the canonical
momentum   ${{\pi}_\mu}^c = \kappa \sqrt{-\gam} \, \p^c X_\mu$ derived from the action
(\ref{2.1}) coincides with the canonical momentum $\kappa (-\gam) \p^c X_\mu/k$ derived from the
action (\ref{2.2}). Additionally, we can choose a gauge so that the determinant factorizes according to
\be
   (-\gam) = {\dot X}^2 (-{\bar \gam}),
\lbl{2.4}
\ee
where ${\dot X}^2 \equiv {\dot X}^\mu {\dot X}_\mu$, ${\dot X}^\mu \equiv \p X^\mu/\p \tau$,
and ${\bar \gam} \equiv {\rm det}\, \p_{\bar a}  X^\mu \p_{\bar b} X_\mu$, ${\bar a},{\bar b} = 1,2,...,p$,
the worldsheet parameters being split as $\xi^a = (\tau, \xi^\ba)$.
Then instead of (\ref{2.2}) we have
\be
  I_{\rm Schild} =\frac{\kappa}{2 k} \int \dd \tau \, \dd^p \sigma \, {\dot X}^2 (-{\bar \gam}),
\lbl{2.5}
\ee
which can be written as
\be
  I _{\rm Schild}= \frac{\kappa}{2 k} \int \dd \tau \, \dd^p \sigma \,\dd^p \sigma' \, (-{\bar \gam})\,
\rho_{\mu \nu} (\sigma,\sigma') \,{\dot X}^\mu (\tau,\sigma){\dot X}^\nu (\tau,\sigma'),
\lbl{2.6}
\ee
 which is the $\tau$-integral of a quadratic form in an infinite dimensional space ${\cal M}$ with the metric
\be
  \rho_{\mu \nu} (\sigma,\sigma') = (-{\bar \gam})\, \eta_{\mu \nu} \, \delta(\sigma-\sigma').
\lbl{2.7}
\ee
At this point it is convenient to introduce a compact notation
\be
   {\dot X}^{\mu (\sigma)}(\tau) \equiv {\dot X}^\mu (\tau,\sigma)\, , ~~~~~~
   \rho_{\mu(\sigma)\nu(\sigma)} \equiv \rho_{\mu \nu} (\sigma,\sigma'),
   \lbl{2.8}
\ee
and write\,\ci{PavsicNewBrane}
\be
   I _{\rm Schild}= \frac{\kappa}{2 {k}} \int \dd \tau \, \rho_{\mu(\sigma)\nu(\sigma')}  {\dot X}^{\mu (\sigma)}(\tau)
  {\dot X}^{\nu (\sigma')}(\tau).
\lbl{2.9}
\ee
Here we use the generalization of Einstein's summation convention, so that not only
summation over the repeated indices $\mu,\nu$, but also the integration over the
repeated continuous indices $(\sigma)$, $(\sigma')$ is assumed. Indices are lowered and
raised, respectively, by $\rho_{\mu(\sigma)\nu(\sigma')}$ and its inverse $\rho^{\mu(\sigma)\nu(\sigma')}$.
The infinite dimensional
space ${\cal M}$ is called {\it brane space}, because its points $x^{\mu(\sg)}$ represent
kinematically possible branes\,\ci{PavsicBook,PavsicPortoroz}.

The quadratic form $\rho_{\mu(\sigma)\nu(\sigma')}  {\dot X}^{\mu (\sigma)}(\tau) {\dot X}^{\nu (\sigma')}(\tau)$
is invariant under diffeomorphisms in the brane space ${\cal M}$. A curve in ${\cal M}$
is given by the parametric equation
\be
   x^{\mu(\sg)} = X^{\mu(\sg)} (\tau),
\lbl{2.10}
\ee
where $X^{\mu(\sg)} (\tau)$ are $\tau$-dependent functions. The velocity of a ``point particle"
in ${\cal M}$ is ${\dot X}^{\mu (\sigma)} \equiv \frac{\p X^{\mu(\sg)}}{\p \tau}$.

The canonical momentum belonging to the action (\ref{2.9}) is
\be
  {p}_{\mu(\sigma)} =\frac{\kappa}{k} \, \rho_{\mu(\sigma)\nu(\sigma')} {\dot X}^{\nu(\sigma')}
  = \frac{\kappa}{k} {\dot X}_{\mu (\sigma)} = \frac{\kappa}{k} \,(-{\bar \gam}) {\dot X}_\mu (\sigma).
\lbl{2.11}
\ee
Its contravariant components are
\be
  p^{\mu (\sigma)} = \rho^{\mu(\sigma) \nu(\sigma')} p_{\nu(\sigma')} = \frac{p^\mu (\sigma)}{(-{\bar \gam})},
\lbl{2.12}
\ee
where $p^\mu (\sigma) = \eta^{\mu \nu} p_\nu (\sigma)$.

The canonical momentum associated with the action (\ref{2.5}) is
\be
  {p}_\mu (\sigma) = \frac{\kappa (-{\bar \gam}) {\dot X}_\mu} {k} = \frac{\kappa \sqrt{-{\bar \gam}}  {\dot X}_\mu} { \sqrt{{\dot X}^\mu {\dot X}_\mu}},
 \lbl{2.13}
\ee
where we have taken into account
\be
  k = \sqrt{{\dot X}^2} \sqrt{-{\bar \gam}} ,
\lbl{2.14}
\ee
which follows from (\ref{2.3a}) and (\ref{2.4}).  Using the latter equation (\ref{2.14}), we verify,
that both momenta, (\ref{2.11}) and (\ref{2.13}), are equal, as they should be. In our notation
$p_{\mu (\sigma)} = p_\mu (\sg)$, whereas $p^{\mu (\sigma)}$ is given by Eq.\,(\ref{2.12}).

The momentum $p_\mu (\sg)$ satisfies the following constraint:
\be
  p_\mu (\sigma) p^\mu (\sigma) = \eta^{\mu \nu} p_\mu (\sg) p_\nu (\sg) = \kappa^2 (-{\bar \gam}).
\lbl{2.15}
\ee
We can also form the quadratic form of the momenta in ${\cal M}$-space,
\be
  p_{\mu(\sigma)} p^{\mu(\sigma)} = \rho^{\mu(\sg) \nu(\sg')} p_{\mu(\sg)} p_{\nu(\sg')} =  {\tl \kappa}^2,
\lbl{2.16}
\ee
in which the integration over repeated indices $(\sg)$ and $(\sg')$ is assumed. Comparing (\ref{2.16})
and (\ref{2.15}), we obtain
\be
  {\tl \kappa}^2 = \int \kappa^2\, \dd \sigma .
\lbl{2.17}
\ee
Let us introduce the quantity
\be
  {\tl k}^2 = \rho_{\mu(\sigma) \nu(\sigma')} {\dot X}^{\mu(\sigma)} {\dot X}^{\nu(\sigma')} =
  \int \dd \sigma (-{\bar \gam}) {\dot X}^\mu {\dot X}^\nu = \int k^2 \dd \sigma ,
\lbl{2.18}
\ee
and take into account that Eqs.\,(\ref{2.17}) and (\ref{2.18}) imply ${\tl \kappa}/{\tl k}=\kappa/k$, i.e.,
\be
  \frac{\tl \kappa}{\sqrt{{\dot X}^{\mu(\sigma)} {\dot X}_{\mu(\sigma)}}} = 
  \frac{\kappa}{\sqrt{(-{\bar \gam}) {\dot X}^\mu {\dot X}_\mu}}.
\lbl{2.19}
\ee
Then we can write the Schild action (\ref{2.9}) in terms of the quantities ${\tl \kappa}$ and ${\tl k}$:
\be
   I _{\rm Schild}= \frac{\tl \kappa}{2 {\tl k}} \int \dd \tau \, \rho_{\mu(\sigma)\nu(\sigma')}  {\dot X}^{\mu (\sigma)}(\tau) {\dot X}^{\nu (\sigma')}(\tau).
\lbl{2.20}
\ee 
The latter action is just a gauge fixed action obtained from the action\,\ci{PavsicBook,PavsicPortoroz,PavsicNewBrane}
\be
   I = {\tl \kappa} \int \dd \tau  \left ( \rho_{\mu(\sigma)\nu(\sigma)}  {\dot X}^{\mu (\sigma)}(\tau)
  {\dot X}^{\nu (\sigma')}(\tau) \right )^{1/2},
\lbl{2.21}
\ee
which gives a minimal length worldline, i.e., a geodesic in ${\cal M}$-space. Indeed, the equations
of motion derived from the action are\,\ci{PavsicNewBrane}
\be
 \frac{\p I}{\p X^{\mu(\sigma)}}= {\tl \kappa} \frac{\dd}{\dd \tau}
  \left ( \frac{{\dot X}_{\mu(\sigma)}}{\sqrt{{\dot {\tl X}}^2}} \right ) -\frac{{\tl \kappa}}{2}\,
  \p_{\mu (\sigma)}  \rho_{\alpha(\sigma')\beta(\sigma'')}\,
  \frac{{\dot X}^{\alpha(\sigma')} {\dot X}^{\beta(\sigma'')}}{\sqrt{{\dot {\tl X}}^2}} = 0, 
\lbl{2.22}
\ee
where
\be
  {\dot {\tl X}}^2 \equiv {\dot X}^{\mu(\sigma)} {\dot X}_{\mu(\sigma)} =
  \rho_{\mu(\sigma) \nu(\sigma')} {\dot X}^{\mu(\sigma)} {\dot X}^{\nu(\sigma')}.
\lbl{2.22a}
\ee 
We use the following notation for functional derivatives:
\be
  \p_{\mu(\sigma)} \equiv \; \, ,_{\mu(\sigma)} \equiv \frac{\p}{\p x^{\mu(\sigma)}} \equiv \frac{\delta}{\delta x^\mu (\sigma)}.
\lbl{2.23}
\ee
Using ${\dot X}_{\mu(\sigma)}=\rho_{\mu(\sigma) \nu (\sigma)} {\dot X}^{\nu (\sigma)}$, and
introducing the connection in ${\cal M}$,
\be
   \Gamma_{\alpha(\sigma') \beta(\sigma'')}^{\,\mu(\sg)}
   = \frac{1}{2}\rho^{\mu(\sigma) \gamma(\sigma''')}
   (\rho_{\gamma(\sigma''') \alpha(\sigma'),\beta(\sigma'')} + \rho_{\gamma(\sigma''') \beta(\sigma''),
   \alpha(\sigma')} - \rho_{\alpha(\sigma') \beta(\sigma''),\gamma(\sigma''')} ),
\lbl{2.24}
\ee
we can write Eq.\,(\ref{2.22}) in the form
\be
  \frac{1}{\sqrt{{\dot {\tl X}}^2}}  \frac{\dd}{\dd \tau}
  \left ( \frac{{\dot X}^{\mu(\sigma)}}{\sqrt{{\dot {\tl X}}^2}} \right )
  + \frac{\Gamma_{\alpha(\sigma') \beta(\sigma'')}^{\mu(\sigma)}
   {\dot X}^{\alpha(\sigma')} {\dot X}^{\beta(\sigma'')}}{{\dot {\tl X}}^2} = 0,
\lbl{2.25}
\ee
which is the equation of geodesic in ${\cal M}$. If we insert into the latter equation the
metric (\ref{2.7}), then we obtain the equation of motion for the Dirac-Nambu-Goto brane.
Equivalently, if we insert the metric (\ref{2.7}) into the action (\ref{2.21}) we obtain
\be
  I= {\tl \kappa} \int \dd \tau \, {\cal L}[{\dot X}^\mu (\sigma), X^\mu (\sigma)],
\lbl{2.26}
\ee
where the Lagrangian
\be
  {\cal L}[{\dot X}^\mu (\sigma), X^\mu (\sigma)]  = \left ( \int \dd^p \sigma\, (-{\bar \gam}) {\dot X}^2 \right )^{1/2}
\lbl{2.27}
\ee
is a functional of infinite dimensional velocities and coordinates. From the Euler-Lagrange equations
\be
   \frac{\dd}{\dd \tau} \frac{\delta {\cal L}}{\delta {\dot X}^\mu (\sigma)} - \frac{\delta {\cal L}}{\delta X^\mu (\sigma)} =0
\lbl{2.28}
\ee
we obtain
\be
   \frac{\dd }{\dd \tau} \left ( \frac{\tl \kappa}{\sqrt{{\dot {\tl X}}^2}} (-{\bar \gam}) {\dot X}_\mu \right )
  + \p_{\bar a} \left ( \frac{{\tl \kappa} (-{\bar \gam}) {\dot X}^2 \p^{\bar a} X_\mu}{\sqrt{{\dot {\tl X}}^2}} \right ) = 0,
\lbl{2.29}
\ee
where ${\dot {\tl X}}^2 \equiv {\dot X}^{\mu(\sigma)} {\dot X}_{\mu(\sigma)} =\int \dd^p \sigma\, (-{\bar \gam}) {\dot X}^2$.
Inserting Eq.\,(\ref{2.19}) into the latter equation, we obtain
\be
   \frac{\dd }{\dd \tau} \left ( \frac{\kappa \sqrt{-{\bar \gam}}}{\sqrt{{\dot X}^2}} {\dot X}_\mu \right )
  + \p_{\bar a} \left ( \kappa \sqrt{-{\bar \gam}} \sqrt{{\dot X}^2} \p^{\bar a} X_\mu \right ) = 0.
\lbl{2.30}
\ee
The same equation follows from the Dirac-Nambu-Goto action (\ref{2.1}) in a gauge (\ref{2.4}).

We have arrived at the minimal length action (\ref{2.21}) by using a particular metric (\ref{2.7}).
However, once we have such an action, we can assume that the metric need not be of that
particular form. We can generalize the validity of the action  (\ref{2.21}) and the corresponding
geodesic equation to any metric. In fact, we can assume that the metric of ${\cal M}$ is
dynamical, like in general relativity, and that to the action (\ref{2.21}) we have to add
a kinetic term for the metric $\rho_{\mu(\sg) \nu(\sg')}$. An approach along such lines was
investigated in Ref.\,\ci{PavsicBook}. Within such a generalized theory the  metric (\ref{2.7}),
leading to the usual Dirac-Nambu-Goto brane is just one of many other possible metrics,
including the metric that is the brane space analog of the flat spacetime metric $\eta_{\mu \nu}$.

\section{Special case: flat brane space ${\cal M}$}

The brane theory simplifies significantly if into the action (\ref{2.21}) we plug the
metric
\be
   \rho_{\mu(\sigma) \nu(\sigma')} = \eta_{\mu(\sigma) \nu(\sigma')} 
   =\eta_ {\mu \nu} \delta (\sigma-\sigma').
\lbl{3.1}
\ee
Then we have\,\ci{PavsicNewBrane}
\be
  I = {\tl \kappa} \int \dd \tau \, \left ( \int \dd^p \, \sigma \, \eta_{\mu \nu} 
    {\dot X}^\mu (\tau,\sigma) {\dot X}^\nu (\tau,\sigma) \right )^{1/2} .
\lbl{3.2}
\ee
This is like an action for a point particle in a flat background space,
\be
  I = m \int \dd \tau \, (\eta_{\mu\nu} {\dot X}^\mu {\dot X}^\nu )^{1/2} ,
\lbl{3.2a}
\ee
\setlength{\unitlength}{.8mm}

\begin{figure}[h!]
\hs{3mm} \begin{picture}(120,125)(0,0)
\put(10,67){\includegraphics[scale=0.35]{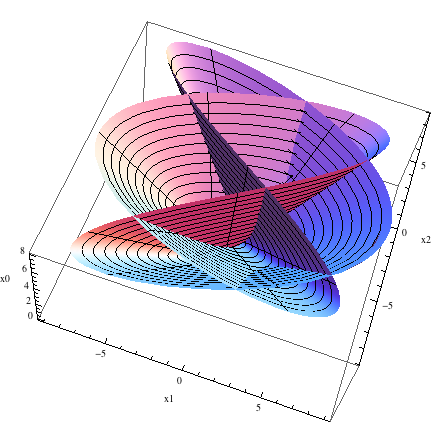}}
\put(100,70){\includegraphics[scale=0.50]{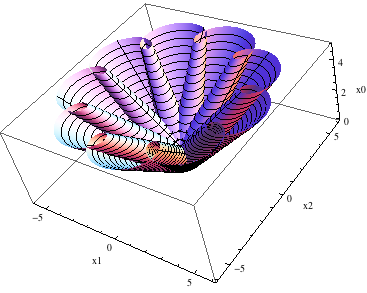}}
\put(5,0){\includegraphics[scale=0.50]{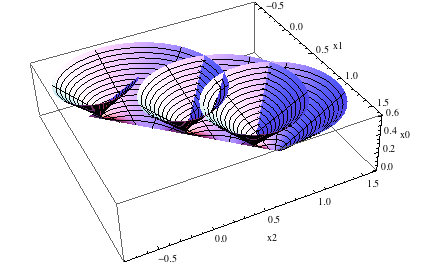}}
\put(105,-5){\includegraphics[scale=0.40]{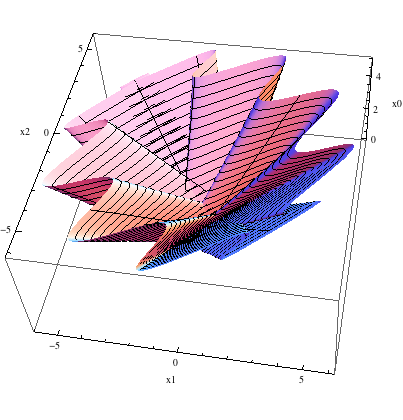}}


\end{picture}

\caption{\footnotesize Examples of flat 1-branes for various choices of initial conditions.}
\end{figure} 
\nnn but the background space is now infinite dimensional. Variation of (\ref{3.2}) gives
the following equations of motion
\be
  \frac{\dd}{\dd \tau} \left ( \frac{{\dot X}^\mu (\tau,\sigma)}{{\sqrt{{\dot{\tl X}}^2}}}
  \right ) = 0,
\lbl{3.3}
\ee
where now we have ${\dot{\tl X}}^2 \equiv   {\dot X}^{\nu(\sigma)} {\dot X}_{\nu(\sigma)} =
\int \dd^p \sigma \, {\dot X}^\mu (\sigma) {\dot X}^\nu (\sigma) \eta_{\mu \nu}$. Choosing
a gauge in which ${\tl {\dot X}}^2 = 1$, we obtain the following simple equations of motion:
\be
   {\ddot X}^\mu (\tau,\sigma) = 0,
\lbl{3.4}
\ee
whose solution is
\be
  X^\mu (\tau,\sigma) = v^\mu (\sigma) \tau + X_0^\mu (\sigma).
\lbl{3.5}
\ee

This describes a bunch of straight worldlines that altogether form a special kind of
brane's worldsheet, namely a worldsheet of a ``flat brane".  Equation\,(\ref{3.5})
thus describes a continuum limit of a system of non-interacting point
particles, tracing straight worldlines.

In Fig.\,2 we give examples of flat 1-branes (i.e., strings) for various solutions
of Eq.\,(\ref{3.5}), i.e., for various choices of $v^\mu (\sigma)$.
In Fig.\,3 we illustrate how the situation looks in the case of a metric that differs from (\ref{3.1}).
For comparison, in Fig.\,4 we show two examples of the usual Nambu-Goto strings.

\setlength{\unitlength}{.8mm}

\begin{figure}[h!]
\hs{3mm} \begin{picture}(120,100)(0,0)
\put(5,0){\includegraphics[scale=0.45]{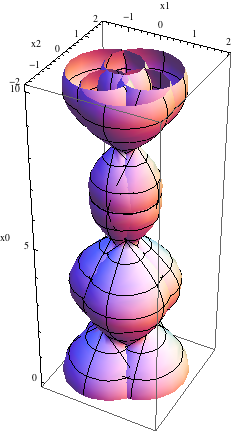}}
\put(60,0){\includegraphics[scale=0.44]{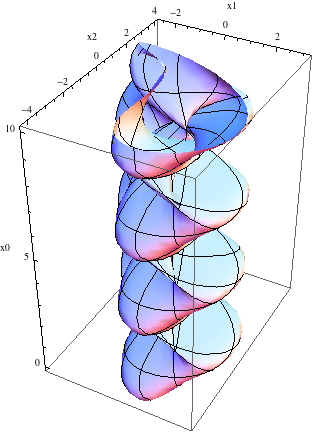}}
\put(125,0){\includegraphics[angle=90,scale=0.50]{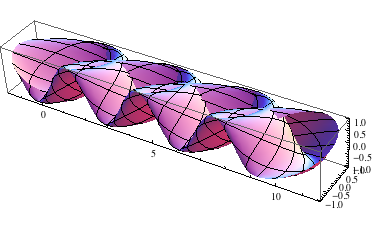}}


\end{picture}

\caption{\footnotesize Examples of ``curved'' 1-branes for various choices of initial conditions.
In all cases the brane space metric is  $\rho_{\mu(\sg) \nu (\sg')} 
= \left (1 + \frac{X'^2 }{{\dot X}^2} \right ) \eta_{\mu \nu} \delta (\sg - \sg')$.}
\end{figure}

We see that flat branes can form involved self intersecting
objects in spacetime. In the last example in Fig.\,2 the worldsheet
does not self intersect, which is a consequence of suitable
boundary conditions.
\setlength{\unitlength}{.8mm}

\begin{figure}[h!]
\hs{3mm} \begin{picture}(120,78)(0,0)
\put(25,0){\includegraphics[angle=90,scale=0.45]{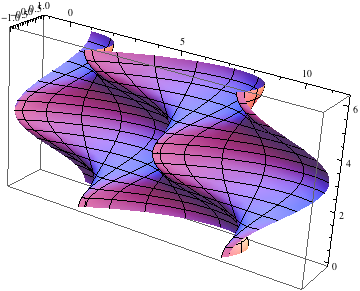}}
\put(95,0){\includegraphics[angle=90,scale=0.50]{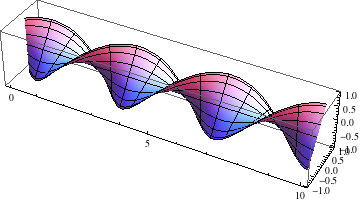}}

\end{picture}

\caption{\footnotesize Examples of a special kind of curved 1-branes: the Nambu-Goto
strings}

\end{figure}

Quantization of the system described by  the action (\ref{3.2}) can be performed
in analogous way as the quantization of the point particle, described
by (\ref{3.2a}).

In the case of the point particle (\ref{3.2a}), we have the constraint
\be
  p^\mu p_\mu - m^2 = 0~,~~~~~~p_\mu = \frac{m {\dot X}_\mu}{({\dot X}^2)^{1/2}},
\lbl{3.6}
\ee
which upon quantization becomes the Klein-Gordon equation,
\be
  \hs{1cm}({\hat p}^\mu {\hat p}_\mu - m^2) \phi (x^\mu) = 0~,~~~~~{\hat p}_\mu
  = - i \p_\mu \equiv - i \frac{\p}{\p x^\mu}.
\lbl{3.7}
\ee

In the case of the brane (\ref{2.21}) with the metric (\ref{3.1}) we have the
constraint
\be
 \hs{1.5cm} p^{\mu(\sigma)} p_{\mu (\sigma)} - {\tl \kappa}^2 = 0~,
  ~~~~~~~p_{\mu(\sigma)} = \frac{{\tl \kappa} {\dot X}_{\mu(\sigma)}}{\sqrt{{\dot{\tl X}}^2}},
\lbl{3.8}
\ee
which upon quantization becomes the generalized Klein-Gordon equation,
\be
   ({\hat p}^{\mu(\sigma)} {\hat p}_{\mu(\sigma)} - {\tl \kappa}^2 ) \phi (x^{\nu(\sigma)}) = 0~,
\lbl{3.8a}
 \ee
\be
      ~~~~~~~{\hat p}_{\mu(\sigma)} \equiv -i \p_{\mu(\sigma)}
   \equiv -i \frac{\p}{\p x^{\mu(\sigma)}}= -i \frac{\delta}{\delta x^\mu (\sigma)}.
\lbl{3.9}
\ee
Here
\be
   p^{\mu(\sigma)} p_{\mu(\sigma)} = \rho_{\mu(\sigma)\nu(\sigma')} p^{\mu(\sigma)} p^{\nu(\sigma')}
   = \int \dd^p \sigma\, \eta_{\mu \nu} p^\mu (\sigma) p^\nu (\sigma),
\lbl{3.10}
\ee
and
\be
  \phi(x^{\nu(\sigma)}) \equiv \phi[x^\mu (\sigma)]
\lbl{3.11}
\ee
is a functional of the brane's embedding functions.
The point particle equation (\ref{3.7}) can be derived from the action
\be
  I[\phi (x^\mu)] = \frac{1}{2} \int \dd^4 x\, (\p_\mu \phi \p^\mu \phi - m^2 \phi^2 ),
\lbl{3.12}
\ee
whereas the corresponding action for the brane equation (\ref{3.8a}) is
\be
  I[\phi(x^{\mu(\sigma)})] = \frac{1}{2} \int {\cal D} x^{\nu(\sigma)} 
  (\p_{\mu(\sigma)} \phi \, \p^{\mu(\sigma)} \phi - {\tl \kappa}^2 \phi^2).
\lbl{3.13}
\ee
Explicitly, the equation of motion derived from the latter action is
\be
   \left ( \p_{\mu(\sigma)} \p^{\mu (\sigma)} + {\tl \kappa}^2 \right ) \phi = 0,
\lbl{3.14}
\ee
In ordinary notation  this reads
\be
 \int \dd^p \sigma \dd^p \sigma' \, \eta^{\mu \nu} \delta (\sigma-\sigma') \,
 \left ( \frac{\delta^2}{\delta x^\mu (\sigma) \delta x^\nu (\sigma')} + {\tl \kappa}^2
 \right ) \phi = 0.
\lbl{3.15}
\ee

As a classical flat brane is like a bunch of free point particles, so a (``first") quantized
brane is like a ``bunch", that is, a continuous set of ``free", i.e., non interacting
quantum fields. Therefore, we can write a solution of Eq.\,(\ref{3.14})
as the product\,\ci{PavsicNewBrane}
\be
  \phi(x^{\mu(\sg)}) = \prod_{\sg''} \vphi^{(\sg'')} (x^{\mu (\sg'')}),
\lbl{3.16}
\ee
where for every  $\sg''$ we have a field $\vphi^{(\sg'')}$ which is a function
of {\it four} spacetime coordinates $x^{\mu (\sg'')}$ that bear a label $\sg''$.
This is just like a separation of variables that is commonly used in solutions
of partial differential equations. We will now use Eq.\,(\ref{2.17}) and introduce
the mass
\be
  m=\kappa \Delta \sg
\lbl{3.17}
\ee
within a region $\Delta \sg \equiv  \Delta^p \sg \equiv \Delta \sg^1 \Delta \sg^2...\Delta \sg^p$.
We will also use the following relation between the functional derivative and
the partial derivative at a fixed point $\sg$ on the brane:
\be
  \p_{\mu(\sg)} \phi \equiv \frac{\delta }{\delta x^\mu (\sg)}\phi = \lim_{\Delta \sg \to 0} \frac{1}{\Delta \sg} \frac{\p_\mu \varphi^{(\sg)}}{\p x^{\mu(\sg)}}  \prod_{\sg'' \neq \sg} \varphi^{(\sg'')} (x^{\mu(\sg'')}).
 \lbl{3.18}
\ee
From (\ref{3.14}),(\ref{3.16})--(\ref{3.18}) we thus obtain
\be
  \left ( \eta^{\mu \nu} \frac{\p^2}{\p x^{\mu (\sg)} \p x^{\nu (\sg)}}+ m^2 \right )
  \vphi^{(\sg)} (x^{\mu(\sg)})= 0.
\lbl{3.19}
\ee
Because $\sg$ is fixed,
we can now rename the four spacetime coordinates $x^{\mu(\sg)}$ into $x^\mu$ and write the latter
equation simply as
\be
  \left ( \eta^{\mu \nu} \frac{\p^2}{\p x^{\mu} \p x^{\nu}}+ m^2 \right )
  \vphi^{(\sg)} (x^{\mu})= 0.
\lbl{3.20}
\ee
In our setup a segment of a classical flat brane around $\sg$ behaves as a free point
particle, and after quantization it satisfies at each $\sg$ the Klein-Gordon equation.
Because $\sg$ is any point on the brane, we have a continuous set of non interacting
scalars fields  $\vphi^{(\sg)}$, every one of them satisfying the Klein-Gordon
equation (\ref{3.20}). In other words,  we describe the flat brane by means of many particle non interacting field theory. Different segments of the brane behave as distinguishable particles, each being described by a different scalar field.

In the case of a discrete set of non interacting scalar fields $\vphi^r (x)$, the system is described by the action
\be
   I[\varphi^r (x)] = \frac{1}{2} \int \dd^D x  \sum_{r=1}^N \Bigl (\p_\mu \vphi^r \p^\mu \vphi^r
   - m^2 (\vphi^r)^2 \Bigr ).
 \lbl{3.21}
\ee

In the continuum limit, the discrete index $r$ becomes the continuous index $(\sg)$ , and $\vphi^r (x)$
becomes $\vphi^{(\sg)}(x) \equiv \vphi(\sg,x)$, or shortly, $\vphi^{(\sg)} \equiv \vphi(\sg)$. The action is then
\be
  I[\vphi{(\sigma)}] = \frac{1}{2} \int \dd^D x \, \int \dd^p \sg \Bigl( \p_\mu \vphi{(\sigma)} \p^\mu \vphi{(\sigma})  - m^2 \vphi^2 {(\sigma)}  \Bigr).
\lbl{3.22}
\ee

A discrete system based on the action (\ref{3.21}) can be straightforwardly second quantized,
and so can be the continuous system (\ref{3.22}). In the discrete case, the canonically conjugate variables, the fields $\vphi^r (t, \bx)$ and momenta $\Pi^r (t,\bx)$, become the operators
satisfying equal $t$ commutation relations
\bear
   &&[\vphi^r(t,{\bx}),\Pi^s(t,{\bx}')] = i \delta^3 ({\bx}-{\bx'}) \delta^{rs}, \nonumber \\
   &&[\vphi^r(t,{\bx}),\vphi^s(t,{\bx'})] =0, ~~~~~[\Pi^r(t,{\bx}),\Pi^s(t,{\bx'})] =0.
\lbl{3.23}
\ear

\section{An interacting bunch of scalar fields}

The action for our system of a continuous set of non interacting scalar fields (\ref{3.22}) can
be written in the form\,\ci{PavsicNewBrane}
\be
  I[\vphi^{(\sigma)}] = \frac{1}{2} \int \dd^D x \, \left ( \p_\mu \vphi^{(\sigma)} \p^\mu \vphi^{(\sigma')}
  - m^2 \vphi^{\sigma)} \vphi^{(\sigma')} \right ) s_{(\sigma)(\sigma')},
\lbl{4.1}
\ee
where
\be
   s_{(\sigma)(\sigma')} = \delta_{(\sg)(\sg')} = \delta (\sg-\sg').
\lbl{4.2}
\ee
The latter form of the action suggests its generalization to a continuous set of
{\it interacting} fields. We see that $s_{(\sigma)(\sigma')}$ has the r\^ole of a
metric in the space of the fields $\vphi^{(\sg)}$. In principle it need not be
the simple metric (\ref{4.2}), but can be a generic metric. In such a way we
introduce interactions among the fields, satisfying the action principle (\ref{4.1})
in which now $s_{(\sigma)(\sigma')}$ is no longer the simple metric (\ref{4.2}),
but a more general metric. 

The equation of motion derived from the action (\ref{4.1}) are
\be
  \p_\mu \p^\mu \vphi_{(\sigma)} + m^2 \vphi_{(\sigma)} = 0,
\lbl{4.3}
\ee
where $\vphi_{(\sg)} = s_{(\sigma)(\sigma')} \vphi^{(\sigma')}$. Assuming that
$s_{(\sigma)(\sigma')}$ has the inverse $s^{(\sigma)(\sigma')}$, so that 
\be
  s^{(\sigma)(\sigma'')}s_{(\sigma'')(\sigma')} = {\delta^{(\sigma)}}_{(\sigma')} \equiv
  \delta (\sigma - \sigma'),
\lbl{4.4}
\ee
then we also have $\vphi^{(\sg)} = s^{(\sigma)(\sigma')} \vphi_{(\sigma')}$.
Applying the latter relation on Eq.\,(\ref{4.3}), we obtain
\be
  \p_\mu \p^\mu \vphi^{(\sigma)} + m^2 \vphi^{(\sigma)} = 0,
\lbl{4.5}
\ee
which is the equation of motion for $\vphi^{(\sg)}$.

A peculiar property of the system so constructed is that even when the
metric is non trivial so that there are interactions among the fields,
a general solution of the equation of motion (\ref{4.3}) has the familiar form
\be
  \vphi_{(\sg)} (x) = \int \frac{\dd^\bD \bk} {\sqrt{(2 \pi)^\bD 2 \omega_\bk}}
  \left ( a_{(\sg)} (\bk) {\rm e}^{-i k x} + a_{(\sg)}^\dg  (\bk) {\rm e}^{i k x} \right ) ,
\lbl{4.6}
\ee
where $\omega_\bk = \sqrt{\bk^2 + m^2}$. The quantities $\vphi_{(\sg)} (x)$,
$ a_{(\sg)} (\bk)$, and  $a_{(\sg)}^\dg  (\bk)$  can be raised by means of the
inverse metric $s^{(\sigma)(\sigma')}$, so that we obtain
\be
  \vphi^{(\sg)} (x) = \int \frac{\dd^\bD \bk} {\sqrt{(2 \pi)^\bD 2 \omega_\bk}}
  \left ( a^{(\sg)} (\bk) {\rm e}^{-i k x} + {a^{(\sg)}}^\dg  (\bk) {\rm e}^{i k x} \right ) ,
\lbl{4.7}
\ee
which is a solution of Eq.\,(\ref{4.5}).

The canonically conjugated variables $\vphi^{(\sigma)}$ and $ \Pi_{(\sigma)} =
\p {\cal L}/\p {\dot \vphi}^{(\sigma)} = {\dot \vphi}_{(\sigma)}$ satisfy 
\be
  [\vphi^{(\sigma)} (x),\Pi_{(\sigma')} (x')]\Bigl\vert_{x^0=x'^0} = {\delta^{(\sigma)}}_{(\sigma')}
  \delta^\bD (\bx-\bx')
\lbl{4.8}
\ee
\be
  [\vphi^{(\sigma)} (x),\vphi^{(\sigma')} (x')] \Bigl\vert_{x^0=x'^0} =0~,~~~~~~~~~~~
   [\Pi_{(\sigma)} (x),\Pi_{(\sigma')} (x')]\Bigl\vert_{x^0=x'^0} =0 .
\lbl{4.9}
\ee
From those quantities we construct the Hamiltonian as usual,
\be
  H= \int \dd^\bD \bx \, (\Pi_{(\sigma)} {\dot \vphi}^{(\sigma)} - {\cal L} ) =
  \frac{1}{2} \int \dd^\bD \bx \, (\Pi_{\sigma)} \Pi^{(\sigma)} - \p_i \vphi^{(\sigma)} \p^i \vphi_{(\sigma)}
  + m^2 \vphi^{(\sg)} \vphi_{(\sg)}).
\lbl{4.10}
\ee
and rewrite it in terms the operators $ a_{(\sg)} (\bk)$, and  $a_{(\sg)}^\dg  (\bk)$.
From Eqs.\,(\ref{4.6})--(\ref{4.9}) we find that the latter operators must satisfy
\be
  [a^{(\sg)}(\bp), a_{(\sg')}^\dg (\bp')] = {\delta^{(\sg)}}_{(\sg')} \delta^\bD (\bp - \bp').
\lbl{4.11}
\ee
\be
    [a^{(\sg)}(\bp), a_{(\sg')} (\bp')] = 0, ~~~~~~~~~[{a^{(\sg)}}^\dg (\bp), a_{(\sg')}^\dg (\bp')] = 0.
\lbl{4.12}
\ee
The relation (\ref{4.11}) can be written in the following equivalent forms:
\be
  [a_{(\sg)}(\bp), a_{(\sg')}^\dg (\bp')] = s_{(\sg)(\sg')} \delta^\bD (\bp - \bp'),
\lbl{4.13}
\ee
 \be
  [a^{(\sg)}(\bp), a^{\dg (\sg')}(\bp')] = s^{(\sg)(\sg')} \delta^\bD (\bp - \bp').
\lbl{4.14}
\ee
The Hamilton then becomes
\bear
  &&H = \frac{1}{2} \int \dd^\bD \bk \, \omega_\bk \left ( a_{(\sg)}^\dg (\bk) a^{(\sg)}(\bk)
  + a^{(\sg)} (\bk) a_{(\sg)}^\dg (\bk) \right ) \nonumber \\
  &&\hs{5mm}=  \int \dd^\bD \bk \, \omega_\bk \, a_{(\sg)}^\dg (\bk) a^{(\sg)} (\bk)+ H_{\rm z.p.},
\lbl{4.15}
\ear
where $H_{\rm z.p.}$ is the "zero point" Hamiltonian, and
\be
  a_{(\sg)}^\dg (\bk) a^{(\sg)}(\bk) = {a^{(\sg)}}^\dg (\bk) a^{(\sg')} (\bk) s_{(\sg)(\sg')}
   =a_{(\sg)} ^\dg(\bk) a_{(\sg')} (\bk) s^{(\sg)(\sg')}.
\lbl{4.16}
\ee

More generally,  by using the standard field theoretic techniques that involve the Noether
theorem, we obtain the stress-energy tensor
\be
  {T^\mu}_\nu = \frac{\p {\cal L}}{\p \p_\mu \vphi^(\sg)} \p_\nu \vphi^{(\sg)} -
  {\cal L} {\delta^\mu}_\nu.
\lbl{4.17}
\ee
Integrating the latter tensor over a space like hypersurface, we obtain the $D$-momentum
$P_\nu = \int \dd \Sigma_\mu {T^\mu}_\nu$. In the reference frame in which  the hypersurface has
components $\dd \Sigma_\mu = (\dd \Sigma_0,0,0,...,0)$ with $\dd \Sigma_0 = \dd^\bD \bx$,
the zero component of the $D$-momentum is the Hamiltonian (\ref{4.10}), whilst the spatial
components are $P_\bmu = \int \dd^\bD \bx \, {\dot \vphi}_{(\sg)} \p_\bmu \vphi^{(\sg)}$, where
$\bmu = 1,2,...,\bD$. After using the expansion (\ref{4.6}),(\ref{4.7}), we have
\bear
   &&{\hat \bp} = \frac{1}{2} \int \dd^\bD \bk \, \bk \left ( a_{(\sg)}^\dg (\bk) a^{(\sg)}(\bk)
  + a^{(\sg)} (\bk) a_{(\sg)}^\dg (\bk) \right ) \nonumber \\
 &&\hs{5mm} =  \int \dd^\bD \bk \,\bk \, a_{(\sg)}^\dg (\bk) a^{(\sg)} (\bk)+ {\hat \bp}_{\rm z.p.},
 \lbl{A4.12}
\ear
where ${\hat \bp}_{\rm z.p.}$ is the ``zero point'' momentum.

By means of the operators $a_{(\sg)}(\bk)$ and $a_{(\sg)}^\dg (\bk)$ we can construct the
states of our system. Defining the vacuum state according to
\be
   a_{(\sg)}(\bk) \vac = 0,
\lbl{A4.13}
\ee
the states with definite momenta are created by $a_{(\sg)}^\dg (\bk)$,
\be
  |\bk_1 \rangle = a_{(\sg)}^\dg (\bk_1) \vac ~,~~~~ |\bk_1 \bk_2 \rangle
  = a_{(\sg)}^\dg (\bk_1) a_{(\sg)}^\dg (\bk_2)\vac ~, ... \, .
\lbl{A4.14}
\ee
These are basis states, from which we can form various more general states. For instance,
we can form single particle wave packet profile states at every $\sg$, and sum (i.e., integrate)
them over $\sg$:
\be
  |\psi \rangle = \int \dd \bp \, g^{(\sg)} (\bp) a_{(\sg)}^\dg (\bp) \vac,
\lbl{A4.15}
\ee
where
\be
  g^{(\sg)} (\bp) a_{(\sg)} (\bp) =g^{(\sg)} (\bp) a^{\dg(\sg')} (\bp) s_{(\sg)(\sg')}.
\lbl{A4.16}
\ee
The action of an annihilation operator on such a state gives
\be
  a^{(\sg')} (\bp') |\psi \rangle = g^{(\sg')} (\bp') \vac,
\lbl{A4.17}
\ee
so that we have
\be
  \langle \psi| a^{\dg (\sg'')} (\bp'') a^{(\sg')} (\bp') |\psi \rangle =
  g^{*(\sg'')} (\bp'') g^{(\sg')} (\bp') \langle 0| 0 \rangle.
\lbl{4.18}
\ee
We normalize the vacuum according to $\langle 0| 0 \rangle = 1$.

Let us now consider the state which is the product of  "single particle" wave
packet profiles\,\ci{PavsicNewBrane}:
\be
~~~~~~~~~~~~~|\psi \rangle = \prod_\sg \int \dd^\bD \bp_{(\sg)} \, g^{(\sg)} (\bp_{(\sg)}) a_{(\sg)}^\dg (\bp_{(\sg)}) \vac ~~~~~~~~~ \mbox{\rm no integration over $(\sg)$}.
\lbl{4.19}
\ee
The action of an annihilation operator to the latter state gives
\be
  a^{(\sg')} (\bp'_{(\sg')}) |\psi \rangle = \int \dd \bp_{(\sg)} \dd \sg' \delta (\sg'-\sg)
   \delta (\bp'_{(\sg')}-\bp_{(\sg)}) g^{(\sg)} (\bp_{(\sg')})|{\bar \psi} \rangle 
   = g^{(\sg')} (\bp'_{(\sg')} |{\bar \psi} \rangle,
\lbl{4.20}
\ee
where $|\bar \psi \rangle$ is the product of all the single "particle" states. except the one picked up by
$a^{(\sg')} (\bp'_{(\sg')})$:
\be
  |{\bar \psi} \rangle = \left ( \prod_{\sg \neq \sg'} \int \dd \bp_{(\sg)} g^{(\sg)} (\bp_{(\sg)})
  a_{(\sg)}^\dg (\bp_{(\sg)}) \right ) \vac.
\lbl{4.21}
\ee
We thus have
\be
  \langle \psi | a^{\dg (\sg'')}(\bp''_{(\sg'')}) a^{(\sg')} (\bp'_{(\sg')}) |\psi \rangle =
  g^{*(\sg'')} (\bp''_{(\sg'')}) g^{(\sg')} (\bp'_{(\sg')}) \langle {\bar \psi}|{\bar \psi} \rangle,
\lbl{4.22}
\ee
where normalization can be such that $ \langle {\bar \psi}|{\bar \psi} \rangle =1$.

We are now going to calculate how the expectation value of the momentum operator
changes with time. Using the Schr\"odinger equation we obtain\,\ci{PavsicNewBrane}
\be
  \frac{\dd}{\dd t} \langle \psi|\hbp |\psi \rangle = \left ( \frac{\dd}{\dd t} \langle \psi | \right ) \hbp |\psi \rangle
  + \langle \psi |\hbp  \frac{\dd}{\dd t} |\psi \rangle = (-i)\langle \psi |\hbp H - H^\dg \hbp |\psi \rangle.
\lbl{4.23}
\ee
In the above derivation we assumed that the Hamilton operator is not Hermitian. This is the case,
if the mass $m= \kappa \Delta \sg$ depends on position $\sg$ on the brane\footnote
{In the discrete case this is equivalent to every particle (field) having a different mass $m_r$.
In the continuous case this means that the brane's tension $\kappa$ is $\sg$ dependent.},
so that also $\sqrt{m^2 (\sg) + \bk^2} = \omega_\bk (\sg)$ is a function of $\sg$. From the expression
(\ref{4.15}) for the Hamiltonian in which instead of a $\sg$ independent $\omega_\bk$
stays $\omega_\bk (\sg)$, we then find $H^\dg \neq H$.

If we insert into Eq.\,(\ref{4.23}) either a state (\ref{4.15}) or (\ref{4.19}) we obtain\,\ci{PavsicNewBrane}
\be
   \frac{\dd}{\dd t} \langle \psi|\hbp |\psi \rangle = (-i) \int \dd^\bD \bp \, \bp \, g^* (\sg,\bp) g(\sg',\bp)
   s(\sg,\sg') (\omega_\bp (\sg) - \omega_\bp (\sg')) \dd \sg \dd \sg' ,
\lbl{4.26a}
\ee
where we now write $g^{(\sg)} (\bp) \equiv g(\sigma,\bp)$, and $s_{(\sg)(\sg')} \equiv s(\sg,\sg')$
and explicitly denote the integration over $\sg$ and $\sg'$
In the case of a $\sg$ independent $\omega_\bp$ the above expression vanishes, which means
that the expectation values of  the system's total momentum is conserved in time. This is
indeed the case for an isolated system, whose tension $\kappa$, and thus $\omega_\bk$
cannot change with $\sg$. If the system is in interaction with another system, then in principle
tension can depend on $\sg$.

Let us now assume that there is the following local interaction between nearby
brane segments\,\ci{PavsicNewBrane}:
\be
  s_{(\sg)(\sg')} \equiv s(\sg,\sg') = (1 + \lambda \p_\ba \p^\ba)\, \delta^p (\sg - \sg').
\lbl{4.27}
\ee
Using the latter expression in Eq.\,(\ref{4.26a}), we obtain
\be
   \frac{\dd}{\dd t}\langle \bp \rangle \equiv  \frac{\dd}{\dd t} \langle \psi|\hbp |\psi \rangle 
   =(-i) \lambda  \int \dd^\bD \bp \, \dd^p \sg \, \bp \,\omega_\bp (\sg) (g^* \p_\ba \p^\ba g - \p_\ba \p^\ba g^* g ).
\lbl{4.28}
\ee
In the expression for the total brane's momentum,
\bear
  \langle \psi|\hbp |\psi \rangle &=& \int \dd \bp \, \dd \sg \, \dd \sigma' \, \bp \,g^*(\sg,\bp) g(\sg',\bp) s(\sg,\sg')
  \nonumber \\
  &=& \int \dd \bp \, \dd \sg \, \bp (g^* g + g^* \p_\ba \p^\ba g ) 
  = \langle \hbp \rangle = \int \dd \sg \langle \hbp \rangle_\sg
\lbl{4.29}
\ear
there is the integrations over $\dd \sg \equiv \dd^p \sg$. If we omit this integration, then
we have the expected momentum density of a brane's segment:
\be
  \langle \hbp \rangle_\sg = \int \dd \bp  \, \bp (g^* g + g^* \p_\ba \p^\ba g ) = \langle \psi|_\sg \hbp |\psi \rangle_\sg,
\lbl{4.30}
\ee
where
\be
  |\psi \rangle_\sg = \int \dd \bp \, g(\sg,\bp) a^\dg (\sg,\bp) \vac ,
\lbl{4.31}
\ee
is the state of the brane's element at $\sg' \equiv \sg'^\ba$, i.e.,  the state (\ref{4.19}) with
the product over $\sg$ being omitted.

The time derivative of such an expected momentum density is obtained from Eq.\,(\ref{4.28}),
if we omit the integrations over $\dd^p \sg$:
\be
   \frac{\dd}{\dd t}\langle \hbp \rangle_\sg 
   =(-i) \lambda  \int \dd^\bD \bp \,\bp \,\omega_\bp (g^* \p_\ba \p^\ba g - \p_\ba \p^\ba g^* g ).
\lbl{4.31}
\ee
The latter expression can be different from zero even if $\omega_\bp$ does not change with $\sg$.
In fact this is the continuity equation for the current density on the brane, isolated from its
environment.

If, instead a wave packet profile in momentum space, we take a wave packet in coordinate
space, the Fourier transformation being
\be
  g(\sg,\bp) = \frac{1}{(2 \pi)^{\bD/2}} \int {\rm e}^{- i \bp \bx} f(\sg,\bx) \dd \bx,
\lbl{4.32}
\ee
then Eq.\,(\ref{4.31}) becomes
\bear
  &&\frac{\dd}{\dd t}\langle \hbp \rangle_\sg 
  = - \lambda \p_\ba \int \dd^\bD \bx \, \left [ f^* (\sg,\bx) 
  \left ( \nabla \sqrt{m^2 + (-i \nabla)^2)}\, \p^\ba f(\sg,\bx) \right ) \right . \nonumber \\
  &&\hs{4cm} -   \left . \left ( \nabla \sqrt{m^2 + (-i \nabla)^2)}\, \p^\ba f^*(\sg,\bx) \right )  f(\sg,\bx) \right ] .
\lbl{4.33}
\ear
Though we have not explicitly denoted so, wave packet profiles $g(\sg,\bp)$ and $f(\sg,\bx)$
depend on time as well. Therefore a state such as (\ref{4.31}) is time dependent and satisfies
the time dependent Schr\"odinger equation with the Hamilton operator (\ref{4.15}). The wave
packet profile then satisfies\,\ci{RelatSchEquation}--\ci{RelatSchEquation3},\ci{Al-Hashimi}
\be
\sqrt{m^2 + (-i \nabla^2)} \, f = - i \frac{\p}{\p t} f .
\lbl{4.34}
\ee
Using the latter equation, we can express (\ref{4.33}) in terms of the time derivative:
\be
  \frac{\dd}{\dd t}\langle \hbp \rangle_\sg 
  = - \lambda \p_\ba \int \dd^\bD \bx \, \left [ f^* (\sg,\bx) \left ( \nabla (-i)\frac{\p}{\p t}\p^\ba f(\sg,\bx) \right )
  -   \left ( \nabla (-i) \frac{\p}{\p t} \p^\ba f^*(\sg,\bx) \right )  f(\sg,\bx) \right ]
\lbl{4.35}
\ee

If we rewrite Eq.\,(\ref{4.35}) in components,
\be
  \frac{\dd}{\dd t}\langle {\hat p}_\bmu  \rangle_\sg 
    = - \lambda \p_\ba \int \dd^\bD \bx \, \left [ f^* (\sg,\bx) \left ( -i \frac{\p}{\p t} \, \p^\ba \p_\bmu f \right )
    -  \left ( -i \frac{\p}{\p t} \, \p^\ba \p_\bmu f^* \right ) f \right ],
\lbl{4.36}
\ee
where $\nabla \equiv \p_\bmu$, $\bmu = 1,2,...\bD$, then we immediately
recognize that the right-hand side of Eq.\,(\ref{4.36}) is the divergence
of the expectation value of the operator
\be
   {{\hat \pi}^\ba}_{~\bmu} = -i \lambda \frac{\p}{\p t}{\stackrel{\leftrightarrow}{\p}}^{\ba} \p_\bmu
  \equiv -i \lambda  \frac{\p}{\p t} \left ( {\stackrel{\leftarrow}{\p}}^{\,\ba}  
  - {\stackrel{\rightarrow}{\p}}^{\,\ba}  \right ) \p_\bmu,
\lbl{4.37}
\ee

Close to the initial time $t=0$ the solution of Eq.\,(\ref{4.34}) for a minimal wave packet
can be approximated with a Gaussian wave packet if its width is greater than the
Compton wavelength:
\be
  f \approx A {\rm e}^{- \frac{(\bx-{\bar {\bm X}}(\sg))^2}{2 {\tl \sg}_0}} {\rm e}^{i {\bar \bp} \bx} {\rm e}^{i {\bar p}_0 t},
\lbl{4.38}
\ee
where ${\bar {\bm X}}(\sg)$, ${\bar \bp}$ and ${\bar p}_0$ are  the coordinates, momentum and energy of the wave packet center, respectively, whilst $A$ is the normalization constant. 

Inserting the wave packet (\ref{4.38}) into (\ref{4.36}), we
obtain\,\ci{PavsicNewBrane}
\be
  \frac{\langle {\hat p}_\bmu \rangle_\sg}{\dd t}
   = - \lambda \p_\ba \left ( \frac{{\bar p}_0}{\Delta S}  \frac{\p^\ba {\bar X}_\bmu}{\sg_o} \right ) ,
\lbl{4.39}
\ee
where $\Delta S= \int \dd^p \sg$. This is reminiscent of the brane equation of motion (\ref{2.30}).

Let us now consider the following metric in the field space, covariant under reparamerizations
of the brane parameters $\sg^\ba$:
\be
  s(\sg,\sg') = \sqrt{-{\bar \gam} (\sg)}\, \delta^p (\sg-\sg') 
  + \p_\ba \left ( \sqrt{-{\bar \gam} (\sg)} \gam^{\ba \bb} \p_\bb \right )
\delta^{p}(\sg - \sg')
\lbl{5.40}
\ee
where $\gam \equiv {\rm det} \gam_{\ba \bb}$. With such a metric, instead of (\ref{4.39})
we obtain\,\ci{PavsicNewBrane}
\be
  \frac{\langle {\hat p}_\bmu \rangle_\sg}{\dd t}
   = - \lambda \p_\ba \left ( \frac{{\bar p}_0}{\Delta S}  \frac{\sqrt{-{\bar \gam}} \,\gam^{\ba \bb}\p_\bb {\bar X}_\bmu}{\sg_o} \right ) ,
\lbl{4.41}
\ee
where now we have $\Delta S = \int \sqrt{- {\bar \gam} (\sg)}$. The latter equation is in fact the equation
of motion (\ref{2.30}) of a classical Dirac-Nambu-Goto brane if we make the following correspondence:
\be
  \langle {\hat p}_\bmu \rangle_\sg \equiv \langle {\hat \bp} \rangle_\sg \longrightarrow
 p_\bmu (\sg) = \frac{\kappa \sqrt{-{\bar \gam}} {\dot X}_\bmu}{\sqrt{{\dot X}^2}}
\lbl{4.42}
\ee
\be
  \frac{{\bar p}_0}{\Delta S} \gam^{\ba \bb} \p^\bb {\bar X}_\bmu \longrightarrow
  \kappa  \sqrt{-{\bar \gam}} \sqrt{ {\dot X}^2} \p^\ba X_\bmu  = p_0 (\sg) {\dot X}^2 \p^\ba X_\bmu ,
\lbl{4.43}
\ee
and take ${\dot X}^2 = 1-v^2 = 1$. The latter equality holds in a gauge in which $\tau = x^0$,
if $v^2 =0$. Recall that Eq.\,(\ref{4.41}) has been calculated for the wave packet at $t \approx 0$,
therefore  $v^2 =0$ is consistent with vanishing $\langle {\hat p}_\bmu \rangle \propto {\dot {\bar X}_\bmu}$,
$\bmu = 1,2,...,\bD$ at $t \approx 0$.

\section{Generalization to arbitrary configurations}

The exercises  with the brane space were just a tip of an iceberg. Instead of 
one brane, a configuration can consist of many branes, or point particles, or both,
as illustrated in Fig.\,4. The action for such a system is a straightforward
generalization of the brane action (\ref{2.21}) to such an extended configurations space $C$:
\be
  I= {\tl \kappa} \int \dd \tau(\rho_{MN} {\dot X}^M {\dot X}^N)^{1/2}
\lbl{5.1}
\ee
Here we use the  same compact indices $M$, $N$ for coordinates in $C$ in various cases:

\hs{18mm}$M= \mu i$    ~~~~~~~~~~~~many point particles
 
\hs{18mm}$M = \mu (\sg)$  ~~~~~~~~~a single brane
  
\hs{18mm}$M = k \mu(\sg)$~~~~~~~~many branes

\hs{18mm}$M= \mu_1 \mu_2 ...\mu_r$ ~~~oriented $r$-volume associated with a brane,\\

\nnn where $\mu=0,1,2,...,\bD$ denotes coordinates of $D$-dimensional spacetime,
$i=1,2,3,....$ counts different particles, and $k=1,2,3,...$ different branes. The meaning
of the last line will be explained shortly below.

\setlength{\unitlength}{.8mm}

\begin{figure}[h!]
\hs{3mm} \begin{picture}(120,55)(-60,0)
\put(0,0){\includegraphics[scale=0.65]{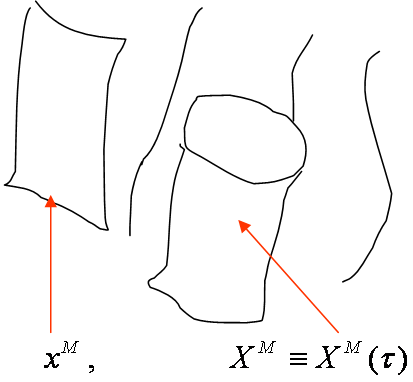}}


\end{picture}

\caption{\footnotesize A configuration can consist of many branes, or point particles, or both.}

\end{figure} 

We thus adopt a generic notation so that $x^M$ and $X^M\equiv X^M (\tau)$
denotes, respectively, coordinates
and $\tau$-dependent functions in whatever configuration space, either a system
of many particles, a single brane, or a system of many branes, or a Clifford
space associated with a brane. Thus, depending on the considered physical
system, $M=\mu i$, $M = \mu(\sigma)$, $M=k \mu (\sigma)$, or
$M= \mu_1 \mu_2 ...\mu_r$. Then Eq.\,(\ref{5.1}) and derived equations are valid
for all those cases of configuration spaces.
\setlength{\unitlength}{.8mm}

\begin{figure}[h!]
\hs{3mm} \begin{picture}(120,50)(0,0)
\put(35,0){\includegraphics[scale=0.65]{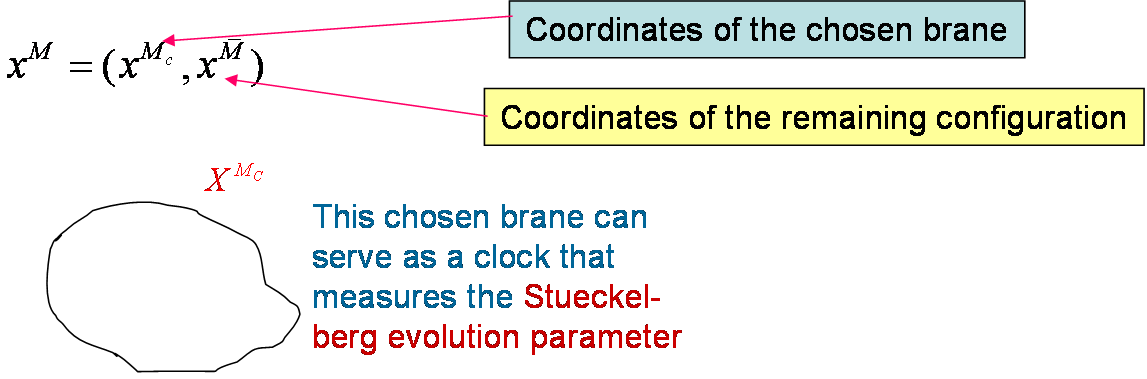}}


\end{picture}

\caption{\footnotesize One of the branes within a configuration can be chosen to serve
as a clock.}

\end{figure}

As a consequence of the invariance of the action (\ref{5.1}) under reparametrizations
of $\tau$, the momenta $p^M = \frac{{\tl \kappa} {\dot X}^M}{\sqrt{{\dot X}^N {\dot X}_N}}$
satisfy the constraint
\be
  p^M p_M - {\tl \kappa}^2 = 0 .
\lbl{5.1a}
\ee

Let us consider a configuration which consists of many particles and/or branes. Let us
choose one brane and denote its coordinates as $X^{M_c}$ (Fig.\,6).

A way to sample a brane is to describe it as a set of 16 oriented $r$-areas (or $r$-volumes)
of all popssible dimensionalities, $r=0,1,2,...,D$. We shall take $D=4$. In Refs.\,\ci{PavsicArena}
it has been shown how a brane, described by an infinite dimensional vector
$x^{\mu(\sigma)}$ can be mapped into a vector of the space spanned by
the basis elements of a Clifford algebra $Cl(1,3)$:
\fontdimen16\textfont2=4pt
\fontdimen17\textfont2=4pt
\be
   x^{\mu(\sigma)} \rightarrow x^{\mu_1 \mu_2 ... \mu_r} \gam_{\mu_a} \wg \gam_{\mu_2}
   \wg ...\wg \gam_{\mu_r} \equiv x^{M_c} \gam_{{M_c}}
\lbl{5.2}
\ee
To avoid multiple counting of the terms, it is convenient to order the indices according
to $\mu_1 <\mu_2 < ...<\mu_r$, $r=0,1,2,3,4$.

\setlength{\unitlength}{.8mm}

\begin{figure}[h!]
\hs{3mm} \begin{picture}(120,20)(-50,0)
\put(0,0){\includegraphics[scale=0.65]{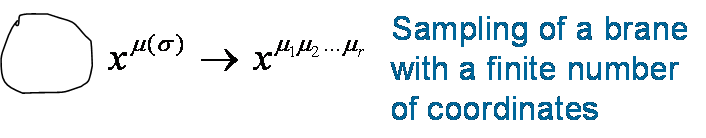}}


\end{picture}

\caption{\footnotesize  A brane can be sampled by coordinates of Clifford space.}
\end{figure} 

If instead of one brane we consider two, three or more branes, such a system
can also be described by 16 coordinates of the Clifford space\,\ci{PavsicArena}
(see Fig.\,8).

\setlength{\unitlength}{.8mm}

\begin{figure}[h!]
\hs{3mm} \begin{picture}(120,60)(-60,0)
\put(0,0){\includegraphics[scale=0.60]{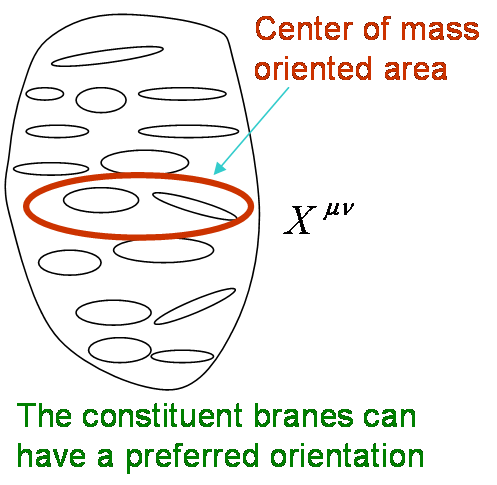}}


\end{picture}

\caption{\footnotesize  An effective (``center of mass'') brane associated with a system
of many branes.}

\end{figure}

Clifford algebra can be considered \ci{Castro1}--\ci{CastroPavsicRev} as a tangent space to a manifold,
called {\it Clifford space}, $C$. We will consider flat Clifford space, which is isomorphic to 
$Cl(1,3)$. Therefore, the points of $C$ can be described by $x^{\mu_1...\mu_r}$.
In eq.\, (\ref{5.2}) we have thus  a mapping from the infinite dimensional brane space
to the 16-dimensional Clifford space. A brane can be sampled by 16 coordinates
$x^M \equiv x^{\mu_1 ...\mu_r}$ of the Clifford space.

The metric of Clifford space is given by the scalar product of two basis elements:
\be
  \eta_{M_c N_c} = \gam_{M_c}^\ddg * \gam_{N_c} 
  = \langle \gam_{M_c}^\ddg \gam_{N_c} \rangle_0,
\lbl{5.3}
\ee
where ``$\ddg$" denotes reversion of the order of vectors in the product
$\gam_{M_c} = \gam_{\mu_1} \gam_{\mu_2} ...\gam_{\mu_r}$.
The subscript ``0" denotes the scalar part of the expression. Explicitly
the metric (\ref{3.7}) is\,\ci{PavsicSaasFee}
\be
  \eta_{M_c N_c} =  {\rm diag} (1,1,1,1,1,1,1,1,-1,-1,-1,-1,-1,-1,-1,-1).
\lbl{5.4}
\ee
Clifford space is thus an ultrahyperbolic space.

The scalar product of $X^\ddg = (x^{M_c} \gam_{M_c})^\ddg$ and
$X = x^{M_c} \gam_{M_c})$ gives
\bear
  X^\ddg * X &=& \eta_{M_c N_c} x^{M_c} x^{N_c} \nonumber\\
    &=& s^2 + \eta_{\mu \nu} x^\mu x^\nu + \mbox{$\frac{1}{4}$}
    (\eta_{\mu\beta} \eta_{\nu \alpha} - \eta_{\mu \alpha} \eta_{\nu \beta})
    x^{\mu \alpha} x^{\nu \beta} + \eta_{\mu \nu} {\tl x}^\mu {\tl x}^\nu
    - {\tl s}^2 \nonumber\\
    &=& \eta_{{\hat \mu}\hnu} x^\hmu x^\hnu +s^2 -{\tl s}^2,
\lbl{5.5}
\ear
where $s$, ${\tl s} = \frac{1}{4!} \epsilon_{\mu \nu \rho \sigma} x^{\mu \nu \rho \sigma}$
and ${\tl x}^\mu = \frac{1}{3!} {\epsilon^\mu}_{\nu \rho \sigma} x^{\nu \rho \sigma}$
are the scalar, pseudoscalar and pseudovector coordinates, respectively.
In the last expression we introduced $x^\hmu = (x^\mu, x^{\mu \nu}, {\tl x}^\mu)$.
We thus have $x^{M_c} = (s,{\tl s},x^{\hat \mu})$.

\fontdimen16\textfont2=2pt
\fontdimen17\textfont2=2pt

Upon (``first") quantization the constraint (\ref{5.1a}), associated with the action (\ref{5.1}),
becomes the Klein-Gordon equation in the configuration space:
\be
  \left ( \rho^{MN} \frac{\p^2}{\p_M \p_N}  + {\tl \kappa}^2 \right ) \phi .
\lbl{5.5b}
\ee
The corresponding action for the scalar field $\phi (x^M)$ is
\be
  I = \int {\cal D} x^M \left ( \rho^{MN} \frac{\p \phi^*}{\p x^{M}} \frac{\p \phi}{\p x^{N}}
    - {\tl \kappa}^2 \phi^* \phi \right ) ,
\lbl{5.5a}
\ee
where ${\cal D}x^M \equiv \underset{M}{\prod} \dd x^M$ is a volume element in the
configuration space. In the case
of many branes, $\underset{M}{\prod} \dd x^M = \underset{k \mu (\sigma)}{\prod} \dd x^{k \mu(\sigma)}$,
whereas in the case of many particles it is 
$\underset{M}{\prod} \dd x^M = \underset{\mu i}{\prod} \dd x^{\mu i}$.

\subsection{Non interacting case}

If the metric $\rho^{MN}$ is a generalization of the Minkowski metric to the configuration space,
then we have the Klein-Gordon equation in flat configuration space. We will now consider
such a non interacting case.

By splitting the index $M$ according to 
$M=(M_c,{\bar M})$, where $M_c$ refers to one chosen brane, described in terms of
the coordinates $x^{M_c} = (s,{\tl s},x^{\hat \mu})$ of the Clifford space, whereas ${\bar M}$
refers to the remaining particle and/or branes, and then renaming ${\bar M}$ back
into $M$, the field action (\ref{5.5a}) becomes
\be
  I = \frac{1}{2} \int {\cal D} x^{M_c}\, {\cal D} x^M \, \left (
  \frac{\p \phi^*}{\p x^{M_c}} \frac{\p \phi}{\p x_{M_c}} +
  \frac{\p \phi^*}{\p x^{M}} \frac{\p \phi}{\p x_{M}} - \kappa^2 \phi^* \phi \right ) ,
\lbl{5.6}
\ee

Let us introduce the light-cone coordinates
\be
  \tau = \frac{1}{\sqrt{2}}(s + {\tl s})~,~~~~~~\lambda = \frac{1}{\sqrt{2}}(s - {\tl s}),
\lbl{5.7}
\ee
so that instead of the coordinates $x^{M_c} = (s,{\tl s}, x^{\hat \mu})$, we have now
the coordinates $x'^{M_c} = (\tau, \lambda, x^{\hat \mu})$.
The field $\phi$ then depends on the light-cone coordinates $\tau$, $\lambda$,
the remaining 12 coordinates  $x^{\hat \mu}$ of the Clifford space associated with
the chosen brane, and on the coordinates $x^M$ of the remaining objects (branes or
particles) within the configurations.

Taking the ansatz
\be
  \phi(\tau,\lambda,x^{\hat \mu},x^M) 
  = {\rm e}^{i \Lambda \lambda} {\rm e}^{i p_{\hat \mu} x^{\hat \mu}} \psi(\tau,x^M),
\lbl{5.7a}
\ee
the action (\ref{5.6}) becomes
\be
  I_0 = \frac{1}{2} \int \dd \tau \, {\cal D} x^M \, \left [
  i \lambda \left ( \frac{\p \psi^*}{\p \tau}\psi - \psi^* \frac{\p \psi}{\p \tau} \right )
  + \p_M \psi^* \p^M \psi - (\kappa^2- p_{\hat \mu} p^{\hat \mu}) \psi^* \psi \right ].
\lbl{5.8}
\ee
We have omitted the integration over $\lambda$ and $x^{\hat \mu}$, because it gives
a constant factor which can be absorbed into the redefinition of the action $I_0$.
The equation of motion is the Stueckelberg equation in the configuration space:
\be
  i \frac{\p \psi}{\p \tau}= - \frac{1}{2 \Lambda} (\p_{M} \p^M  + M'^2) \psi ,
\lbl{5.9}
\ee
where ${M'}^2 = \kappa^2 - p_{\hat \mu} p^{\hat \mu}$. The general solution is
\be
  \psi(\tau,x^M) = \int {\cal D} p^M \, c(p^M) \, {\rm exp} \left [ i p_M x^M
  - \frac{i}{2 \Lambda} (p_M p^M - {M'}^2 ) \tau \right ],
\lbl{5.10}
\ee
in which there is no restriction on momenta $p_M$, therefore initial data at
$\tau=0$ can be freely specified.

In particular it can be  $c(p^M) = c(p^{1 \mu(\sigma)})  c(p^{2 \mu(\sigma)}) ...c(p^{N \mu(\sigma)})$,
for a multi brane configuration, or $c(p^M) = c(p^{1 \mu}) c(p^{2 \mu}) ...c(p^{N \mu})$ for 
a multi particle configuration. Then the field $\phi(x^M)$ can be written as the product of
sigle brane or single particle states. In the case of particles we have:
\be
  \psi (\tau,x^M) = \varphi (\tau,x^{1 \mu}) \varphi (\tau, x^{2 \mu}) ... \varphi (\tau, x^{N \mu})
  \, {\rm e}^{\frac{i}{2 \Lambda} {M'}^2 \tau},
\lbl{5.11}
\ee
where
\be
  \varphi(\tau,x^{1 \mu}) = \int \dd^4 p_1 \, c(p^{1 \mu})\,  {\rm exp} \left [ i p_{1 \mu} x^{1 \mu} -
  \frac{i}{2 \Lambda} \, p_{1 \mu} p^{1 \mu} \tau \right ],
\lbl{5.11a}
\ee
and similarly for other particles labelled by $2,3,...,N$.

Writing now
\be
  \psi(\tau, x^M) = \varphi (\tau, x^\mu) \chi (x^{\bar M}),
\lbl{5.12}
\ee
where $\varphi (\tau, x^\mu) \equiv \varphi (\tau, x^{1 \mu})$ is the field associated with 
a chosen particle (labelled by `1'),  and
\be
   \chi(\tau, x^{\bar M}) = \int \dd \tau \, \dd^{{\cal D}-4} {\bar p} \, c({\bar p}) \, {\rm e}^{i p_{\bar M} x^{\bar M}}
   {\rm e}^{- \frac{i}{2 \Lambda} (p_{\bar M} p^{\bar M} - M'2 ) \tau} 
\lbl{5.13}
\ee
is the field over the configuration of the remaining particles with coordinates
$x^{\bar M} \equiv x^{{\bar i} \mu}$, ${\bar i} = 2,3,...,N$, the action
(\ref{5.8}) becomes
\be
  I = Q \int  \dd \tau \,\dd^4 x  \left [ - i 2 \Lambda \varphi^* \frac{\p \varphi}{\p \tau}
  + \p_\mu \varphi^* \p^\mu \varphi - m^2 \varphi^* \varphi \right ],
\lbl{5.14}
\ee
where
\be
  m^2 = \frac{(1-Q)}{Q} M'^2,
\lbl{5.14a}
\ee
and
\be
  Q = \int \dd^{{\cal D}-4} {\bar x} \, \chi^* \chi = \int \dd^{D-4} {\bar p} \, c^* ({\bar p}) c({\bar p}).
\lbl{5.15}
\ee
We can normalize $\chi$ so that $Q=1$. Then
\be
  I =  \int \dd^4 x \left [ - i 2 \Lambda \varphi^* \frac{\p \varphi}{\p \tau}
  + \p_\mu \varphi^* \p^\mu \varphi \right ],
\lbl{5.16}
\ee
which is the Stueckelberg action\,\ci{StueckelbergField1}--\ci{StueckelbergField11},\ci{PavsicBook}
for a single particle field $\varphi(\tau,x^\mu)$.
From (\ref{5.16}) we obtain the Stueckelberg field equation
\be
  i \frac{\p \varphi}{\p \tau}= - \frac{1}{2 \Lambda} \p_\mu \p^\mu \varphi.
\lbl{5.17}
\ee
The non interacting many particle Stueckelberg equation (\ref{5.9}) thus contains
the single particle Stueckelberg equation (\ref{5.17}). 

Upon quantization, $\varphi (\tau, x^\mu)$ becomes the operator that annihilates,
and $\varphi^* (\tau, x^\mu)$  the operator $\varphi^\dg (\tau,x^\mu)$
that creates a particle (more precisely,  an `instantonic' particle or and `event') 
at $x^\mu$. The evolution of the system is given in terms of the Stueckelberg
evolution parameter $\tau$, which in our setup is associated with the brane
sampled by the coordinates $x^{M_c}$ of the Clifford space. The latter brane\footnote{
It need not be only one brane, there can be many branes, altogether sampled
by $x^{M_c}$ (see Ref.\,\ci{PavsicArena}).}
is a part of the overall considered configuration, and is given the role of a clock,
which can be a ``Stueckelberg clock". The Stueckelberg evolution parameter $\tau$
is thus embedded in the configuration.

In the Stueckelberg quantum field theory, the position operator is not considered as
problematic\footnote{A careful analysis reveals that position operator is not
problematic\,\ci{PavsicNewBrane} even in the usual quantum field theory.}.
It creates an {\it event} in spacetime.

\subsection{Bunch of Stuckelberg fields interacting in a particular way}

The procedure with branes and interacting quantized fields that we have performed in Sec.\,4
can be done \`a la Stueckelberg as well. The Stueckelberg field action (\ref{5.16}) or its
more general form (\ref{5.14}) refers to a
single quantum field. Instead of one such a field we can have many fields, and even
a continuous set of such fields, as in Sec.\,4. But instead of the field action (\ref{4.1}) we
now have (upon quantization) the following action
\be
  I[\vphi^{(\xi)}] = \frac{1}{2} \int \dd \, \tau \dd^D x 
  \left ( - i 2 \Lambda \vphi^{\dg(\xi)} \frac{\p \vphi^{(\xi')}}{\p \tau}
  +  \p_\mu \vphi^{\dg(\xi)} \p^\mu \vphi^{(\xi')} 
  - m^2 \vphi^{\dg(\xi)} \vphi^{(\xi')}  \right ) s_{(\xi)(\xi')},
\lbl{5.18}
 \ee
 where $\xi \equiv \xi^a$, $a=1,2,...,d$,  are $d$ parameters.
 If the metric is $s_{(\xi)(\xi')}= \delta (\xi,\xi')$, then this is the action for
a continuous set of non interacting Stueckelberg fields, otherwise it is an
action for interacting Stueckelberg fields. The momentum, canonically conjugate
to the field $\vphi^{(\xi)} (x)$ is  $\Pi^{(\xi)}=-i \Lambda \vphi^{\dg (\xi)} (x)$. We have
the following commutation relations
\be
  [\vphi^{(\xi)}(\tau,x) \Pi_{(\xi')} (\tau,x')] = i {\delta^{(\xi)}}_{(\xi')} \delta^D (x-x'),
\lbl{5.19}
\ee
\be
  [\vphi^{(\xi)}(\tau,x),\vphi^{(\xi')}(\tau,x')] = 0~,~~~~~~~ [\Pi^{(\xi)}(\tau,x),\Pi^{(\xi')}(\tau,x')] = 0.
\lbl{5.19a}
\ee

The equation of motion derived from (\ref{5.18}) is
\be
  i \frac{\p \vphi_{(\xi')}}{\p \tau} = - \frac{1}{2 \Lambda} (\p_\mu \p^\mu + m^2) \vphi_{(\xi')}.
\lbl{5.20}
\ee
Its solution can be expanded according to
\be
   \vphi_{(\xi)} (\tau,x)= \frac{1}{\sqrt{(2 \pi)^{D} \Lambda}}\int \dd^D p \, a_{(\xi)}(p) {\rm exp} \left [ i p_\mu x^\mu + \frac{i}{\Lambda} (p^\mu p_\mu - m^2) \tau \right ],
\lbl{5.21}
\ee
where the commutation relations (\ref{5.19}) are satisfied provided that
\be
    [a^{(\xi)}(p), a_{(\xi')}^\dg (p')] = {\delta^{(\xi)}}_{\xi')} \delta^D (p - p'),
\lbl{5.22}
\ee
while, as usually, the commutators of equal type operators, vanish.

An operator $a_{(\xi)}^\dg (p)$ creates and $a^{(\xi)}(p)$ annihilates a $(\xi)$-type
particle with momentum $p \equiv p^\mu$, $\mu = 0,1,2,...,D$.
Vacuum state is defined according to $a^{(\xi)}(p)\vac = 0$. The Fourier transformed
operators
\be
   a_{(\xi)}^\dg (x)= \frac{1}{\sqrt{(2 \pi)^{D}}}\int \dd^D p \, a_{(\xi)}^\dg(p) {\rm e}^{ -i p_\mu x^\mu} ,
 \lbl{5.23}
\ee
\be
   a_{(\xi)} (x)= \frac{1}{\sqrt{(2 \pi)^{D}}}\int \dd^D p \, a_{(\xi)}(p) {\rm e}^{ i p_\mu x^\mu} ,
 \lbl{5.24}
\ee
are creation and annihilation operators for a particle event at a spacetime point $x^\mu$. Up to
a factor $\sqrt{\Lambda}$ they coincide with the field operators  $\vphi_{(\xi)} (\tau,x)$ and
$\vphi_{(\xi)}^\dg (\tau,x)$ at a fixed value of $\tau$ (say $\tau=0$).

A many particle event state is obtained by successive action of creation operators on the vacuum.
In the limit of infinitely many densely packed events such a configuration can be a brane (an extended
event) in spacetime:
\be
  \prod_{\xi} a_{(\xi)}^\dg (x_\xi)\vac  \equiv A^\dg [X^\mu (\xi)] \vac = |X^\mu (\xi)\rangle ,
\lbl{5.25}
\ee
where $X^\mu (\xi)$ are a brane's embedding functions of $d$ parameters $\xi \equiv \xi^a$,
which now need not be all space like; one of them can be time like\,\ci{PavsicBook}. In such a case
$X^\mu (\xi)$ describes a brane that extends into $d-1$ spacelike directions and into one time
like direction of the embedding space. General states are superposition of
the states (\ref{5.25}) or their momentum space counterparts.

The Hamilton operator is
\bear
  &&H = \int \dd^D x (\Pi_{(\xi)} \p_{\tau} \vphi^{(\xi)} - {\cal L} )\nonumber\\
  &&~~~ =\frac{1}{2 \Lambda}
  \int \dd^D x \, (\p_\mu \vphi^{\dg (\xi)} \p^\mu \vphi^{(\xi')} 
  - m^2 \vphi^{\dg (\xi)} \p^\mu \vphi^{(\xi')}) s_{(\xi)(\xi')}
  \nonumber\\
  &&~~~=\frac{1}{2 \Lambda} \int \dd^D p\,( p^2-m^2)\, a^{\dg(\xi)}(p) a^{(\xi')}(p) s_{(\xi)(\xi')}.
\lbl{5.26}
\ear  
Similarly, we obtain the momentum operator:
\be
{\hat p}_\mu = \int \dd^D p\, p_\mu \, a^{\dg(\xi)}(p) a^{(\xi')}(p) s_{(\xi)(\xi')}
\lbl{5.27}
\ee

Let us now calculate how the expectation value of the momentum
operator changes with the evolution parameter $\tau$. The procedure is
analogous to that in Sec.\,4. Instead of the state (\ref{4.19}) we now take
\be
~~~~~~~~~~~~~|\psi \rangle = \prod_\xi \int \dd^D p_{(\xi)} \, g^{(\xi)} (p_{(\xi)}) a_{(\xi)}^\dg (p_{(\xi)}) \vac ~~~~~~~~~ \mbox{\rm no integration over $(\xi)$}.
\lbl{5.28}
\ee
Taking $m = m(\xi)$ and introducing
 \be
  h(p,\xi) = \frac{\Lambda}{2} ( p^2-m^2),
\lbl{5.29}
\ee 
we obtain
\be
   \frac{\dd}{\dd \tau} \langle \psi| {\hat p}_\mu |\psi \rangle = (-i) \int \dd^D p \, p_\mu \, g^* (\xi,p) g(\xi',p)
   s(\xi,\xi') (h(p,\xi) - h(p,\xi')) \dd \xi \dd \xi' ,
\lbl{5.30}
\ee
where $g(\xi,p) \equiv g^{(\xi)} (p)$ and $s(\xi,\xi') \equiv s_{(\xi)(\xi')}$.

If we take  the field space metric
\be
  s(\xi,\xi')= (1 + \lambda_c \p^a \p_a)\, \delta^d (\xi - \xi'),
\lbl{5.31}
\ee
then
\be
   \frac{\dd}{\dd \tau}\langle p_\mu \rangle \equiv  \frac{\dd}{\dd \tau} \langle \psi|{\hat p}_\mu |\psi \rangle 
   =(-i) \lambda_c  \int \dd^D p \, \dd^p \xi \, p_\mu \, h(p,\xi) (g^* \p_a \p^a g - \p_a \p^a g^* g ).
\lbl{5.32}
\ee
This is the time derivative of the expectation value of the total momentum of the brane, and it vanishes
if $h(p,\xi)$ does not change with $\xi$.

The expectation value of the total momentum is given by the integral over the momenta of
the brane's segments:
\bear
  \langle \psi|{\hat p}_\mu |\psi \rangle &=& \int \dd p \, \dd \xi \, \dd \sigma' \, p_\mu \,g^*(\xi,p) g(\xi',p) s(\xi,\xi')
  \nonumber \\
  &=& \int \dd p \, \dd \xi \, p_\mu (g^* g + g^* \p_a \p^a g ) 
  = \langle {\hat p}_\mu \rangle = \int \dd \xi \langle {\hat p}_\mu \rangle_\xi
\lbl{5.33}
\ear
where
\be
  \langle {\hat p}_\mu \rangle_\xi = \int \dd p  \, p_\mu (g^* g + g^* \p_a \p^a g )
   = \langle \psi|_\xi {\hat p}_\mu |\psi \rangle_\xi
\lbl{5.34}
\ee
From Eq.\,(\ref{5.32}) we then read the following expression for the time derivative of 
the momentum of a brane segment:
\be
   \frac{\dd}{\dd \tau}\langle p_\mu \rangle_\xi 
   =(-i) \lambda_c  \int \dd^D p  \, p_\mu \, h(p,\xi) (g^* \p_a \p^a g - \p_a \p^a g^* g ),
\lbl{5.35}
\ee 
which in general is different from zero even if $h(p,\xi)$ does not change with $\xi$.

If in Eq.\,(\ref{5.35}) we express the wave packet profile $g(\xi,p)$ in term of its position
space counter part $f(\xi,p)$,
\be
  g(\xi,p) = \frac{1}{(2 \pi)^{D/2}} \int {\rm e}^{- i p_\mu x^\mu} f(\xi,x) \dd x,
\lbl{5.36}
\ee
then we obtain
\be
  \frac{\dd}{\dd \tau}\langle {\hat p}_\mu  \rangle_\xi 
    = - \lambda_c \p_a \int \dd^D x \, \left [ f^* (\xi,x) \left ( \frac{1}{2 \Lambda}(\p_\mu \p^\mu- m^2) \, \p^a \p_\mu f \right )
    -  \left ( \frac{1}{2 \Lambda}(\p_\mu \p^\mu- m^2)\, \p^a \p_\mu f^* \right ) f \right ].
\lbl{5.37}
\ee

Though not written explicitly, the wave packet profiles $g$ and $f$ depend on the evolution time $\tau$.
Using the Schr\"odinger equation with the Hamiltonina (\ref{5.26}) for the state (\ref{5.28}), we obtain
the equation of motion for the wave packet profile $f$:
\be
\frac{1}{2 \Lambda}(\p_\mu p^\mu- m^2) f = - i \frac{\p}{\p \tau} f.
\lbl{5.38}
\ee
Using the latter equation in Eq.\,(\ref{5.37}) we obtain
\be
  \frac{\dd}{\dd \tau}\langle {\hat p}_\mu  \rangle_\xi 
    = - \lambda_c \p_a \int \dd^D x \, \left [ f^* (\xi,x) \left ( -i \frac{\p}{\p \tau} \, \p^a \p_\mu f \right )
    -  \left ( -i \frac{\p}{\p \tau} \, \p^a \p_\mu f^* \right ) f \right ],
\lbl{5.39}
\ee

For a Gaussian wave packet profile
\be
  f \approx A {\rm e}^{- \frac{(x-{\bar  X}(\xi))^2}{2 {\tl \sg}_0}} {\rm e}^{i {\bar p}_\mu x^\mu} {\rm e}^{i {\bar h}, \tau},
\lbl{5.40}
\ee
where
\be
  {\bar h}=  \frac{1}{2 \Lambda} ( {\bar p}^2-m^2),
\lbl{5.41}
\ee
equation (\ref{5.39}) becomes
\be
  \frac{\langle {\hat p}_\mu \rangle_\xi}{\dd \tau}
   = - \lambda_c \p_a \left ( \frac{{\bar h}}{\Delta S}  \frac{\p^a {\bar X}_\mu}{{\tl \sg}_0} \right ) .
\lbl{5.42}
\ee

The expressions with the metric (\ref{5.31}) are not covariant with respect to arbitrary reparametrizations
of $\xi^a$. If we take the  metric
\be
  s(\xi,\xi') = \sqrt{-\gam(\xi)} \delta^{d} (\xi-\xi') + \p_a \left (\sqrt{-\gam(\xi)} \gam^{ab} \p_b \right ) 
  \delta^{d}(\xi - \xi')
\lbl{5.43}
\ee
where $\gam \equiv {\rm det} \gam_{ab}$, then the expressions become covariant, and
instead of (\ref{5.42}) we obtain
\be
  \frac{\langle {\hat p}_\mu \rangle_\xi}{\dd \tau}
   = - \lambda_c \p_a \left ( \frac{{\bar h}}{\Delta S} 
    \frac{\sqrt{-\gam(\xi)}\gam^{ab}\p_b {\bar X}_\mu}{{\tl \sg}_0} \right ) .
\lbl{5.44}
\ee
The latter equation tells how the expected momentum density $\langle {\hat p}_\mu \rangle_\xi$ changes
with the evolution parameter $\tau$, which in the Stueckelberg theory is the ``true" time, whereas
$x^0 \equiv t$ is just one of spacetime coordinates. 
In Appendix we show that Eq.\,(\ref{5.44})  corresponds to the equation of motion
of a classical Stueckelberg brane (see \ci{PavsicBook}), which is a generalization
of the Stueckelberg point particle.

\subsection{Self interacting Stueckelberg field in configuration space}

In the absence of interactions, a field $\psi(\tau, x^M) \equiv \psi(\tau, x^{1\mu},
x^{2 \mu},...,x^{N \mu})$ over a many particle configuration is the product
(\ref{5.11}) of the single particle fields. In the presence of interactions, in general
this is no longer the case. An interacting field theory is described by the
action (\ref{5.8}) to which we add an interactive term $I_{\rm int}$,
 so that the total action
is
\be
  I = I_0 + I_{\rm int}.
\lbl{5.45}
\ee
We will take $I_{\rm int} = - \frac{G_0}{4!} (\psi^* \psi)^2$.
Let us also assume that a particle, say, No.\,1, can be singled out from the rest
of the configuration according to (\ref{5.12}). Inserting Eq.\,(\ref{5.12})
into the action (\ref{5.45}), we obtain the Stueckelberg action for the scalar
field $\varphi(\tau, x^\mu)$ with the quartic self interaction:
\be
  I =  \int \dd \tau\,\dd^4 x \left [ - i 2 \Lambda \varphi^* \frac{\p \varphi}{\p \tau}
  + \p_\mu \varphi^* \p^\mu \varphi + m_{\rm res} \varphi^* \varphi - g_0 (\varphi^* \varphi)^2  \right ],
\lbl{5.46}
\ee
where $g_0 = G_0 \int \dd^{{D}-4} {\bar x} (\chi^* \chi)^2$, and
\be
   m_{\rm res}^2 = \int \dd^{{\cal D}-4} {\bar x} \, \left ( \p_{\bar M} \chi^* \p^{\bar M} \chi
   - i 2 \Lambda \chi^* \frac{\p \chi}{\p \tau} \right ),
\lbl{5.47}
\ee
is the residual mass
that is determined by the presence of the field $\chi (\tau, x^{\bar M})$ due to all
the other particles of the configuration. In general, $m_{\rm res}^2$ is different
from zero. In particular, in the absence of an interaction,  $\chi$ is given by
Eq.\, (\ref{5.13}) and then $m_{\rm res}^2 =0$.

For an interacting field theory the factorization  (\ref{5.11}) of a field
$\psi(\tau,x^M)$ is valid only if the particle No.\,1 is not entangled with the
other, mutually interacting, particles. If it is entangled, then (\ref{5.11}) does not
hold. We must then work with the field $\psi(\tau,x^M)$ without factoring out
a single particle field.

We have thus arrived at the many particle analog of the brane theory,
described by the classical action (\ref{2.21}) or the first quantized action
(\ref{3.13}), in which now the metric $\rho_{\mu(\sigma) \nu(\sigma')}$ of  the
brane space is not flat. Then one cannot describe a brane as a bunch
of point particles. Similarly, in general one cannot describe a many
particle configuration as a bunch of point particles. Only if the metric
is $g_{MN} \equiv g_{(i \mu)(j \nu)} = \delta_{ij} g_{\mu \nu}$ 
one has a bunch of point particles. In general, the metric need not
be diagonal in the indices $(i \mu)$, $(j \nu)$. Then the particles are
intertwined more than it is usually assumed. The physics, either classical or
quantized, has to be done in a configuration space ${\cal C}$ of many
particles/branes. The metric $\rho_{MN}$ of ${\cal C}$ in general is curved. An interactive
term such as $I_{\rm int} = - \frac{G_0}{4!} (\psi^* \psi)^2$ can be obtained from
the dimensional reduction of the action of the form (\ref{5.8}), along the lines
similar to that of Ref.\,\ci{PavsicFirenze}.

\section{Conclusion}

Within this approach configuration  space ${\cal C}$ is primary even in classical physics,
and the action principle must be formulated in ${\cal C}$, not in spacetime. In other words,
physics, both classical and quantum, must be formulated in configuration space which can be a space
of many point particles and/or branes.
Space or spacetime is a subspace of a configuration space (Fig.\,9). The concept of spacetime has to be revised by considering spacetime as a subspace of a configuration space, which
ultimately is that of the whole universe.

\setlength{\unitlength}{.8mm}

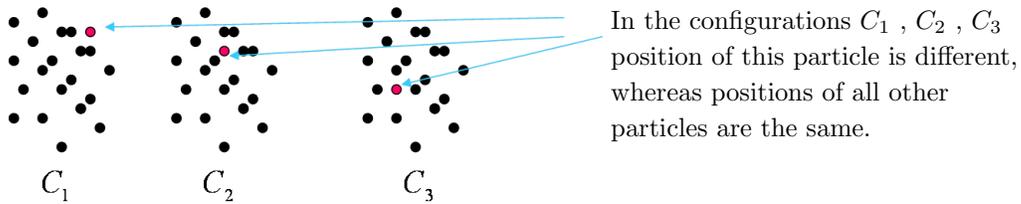
\begin{figure}[h!]
\hs{3mm} \begin{picture}(120,47)(-7,0)
\put(0,0){\includegraphics[scale=0.60]{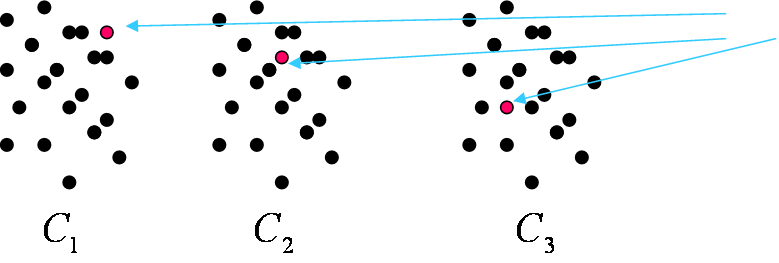}}

\put(100,30){\footnotesize In the configurations $C_1$ , $C_2$ , $C_3$}
\put(100,24){\footnotesize position of this particle is different,}
\put(100,18){\footnotesize whereas positions of all other}
\put(100,12){\footnotesize particles are the same.}

\end{picture}

\caption{\footnotesize  Space(time) as a subspace of configuration space. }
\end{figure}

We have arrived at such conclusion by inspecting the action of a Dirac-Nambu-Goto
brane. We have found that a brane can be considered as a point in an infinite dimensional
brane space ${\cal M}$, moving along a geodesic in ${\cal M}$. The metric of ${\cal M}$ is not fixed, it is dynamical,
like in general relativity. For a particular metric we obtain the usual Dirac-Nambu-Goto brane.
More general metrics give us interesting fancy branes (Fig.\,3) that might be useful in scenarios for
quantum gravity in the presence of matter, where matter is given by the brane's self
intersections\,\ci{PavsicBook,PavsicBrane}.
 The simplest is the ``flat'' metric that gives
us ``flat branes'' (Fig.\,2).  A flat brane can be straightforwardly quantized as a bunch of point particles.
If we take suitable interactions between the quantum fields, we obtain as an
``expectation value'' the classical Dirac-Nambu-Goto brane. 

We have thus found how to quantize branes: via flat brane space. Non flat branes
are then objects of an effective classical theory that arises from the underlying QFT
of many interacting fields.

The concept of configuration space is associated not only with branes, but with
whatever physical systems, in classical and quantum theory. A configuration can be:

- a single brane, considered as a bunch of point particles,

- a discrete system of point particles,

- a mixed system of many branes and point particles,

-  etc.

A closed brane or a system of closed branes (Fig.\,8) can be approximately described
by a finite number of degrees of freedom, which are coordinates of the 16-dimensional
Clifford space. The latter space has signature (8,8), i.e., its points can be described by
eight ``time like'' coordinates
(associated with the plus sign of the metric) and eight ``space like'' coordinates (associated with
the minus sign of the metric). By picking up one time like and one space like coordinate,
and composing from them the analog of two light-cone coordinates, we have derived
the Stueckelberg action for a scalar field. We have also shown how a continuous set
of such locally interacting fields leads to the effective classical branes \`a la Stueckelberg.
The latter objects satisfy the equations of motion that can be obtained by calculating
the time derivative of the expectation value of the momentum operator with respect
to certain ``wave packet" like quantum states created by the Stueckelberg field
operators.

\vs{7mm}

\centerline{\bf Acknowledgement}

This work has been supported by the Slovenian Research Agency.

\vs{1cm}

{\bf \large Appendix: Stueckelberg point particle and its generalization to a brane}

\vs{2mm}

The phase space action for a point particle in a $(D+2)$-dimensional space with signature $(2,D)$ is
\be
  I = \int \dd \tau \left [ p_M {\dot x}^M - \frac{\alpha}{2}(p_M p^M - M^2) \right ],
\lbl{A1}
\ee
where $\alpha$ is a Lagrange multiplier whose variation gives the constrain $p_M p^M - M^2 =0$.
The signature of the extra two dimensions is $(+ -)$, whilst the signature of the $D$-dimensional
space is $(1,D-1)$. Now we take $D=4$, so that we have an extra fifth and sixth dimension.
If $x^5$ and $x^6$ are ``light-cone'' coordinates, then the action reads
\be
  I =  \int \dd \tau \left [ p_\mu {\dot x}^\mu +p_5 {\dot x}^5 +p_6 {\dot x}^6
- \frac{\alpha}{2}(p_\mu p^\mu    - 2 p_5 p_6 - M^2) \right ].
\lbl{A2}
\ee
Taking a gauge in which $\tau = x^5$, we have
\be
  p_6 \equiv - \Lambda = \frac{{\dot x}_6}{ \alpha};~~~~ x^5 = \tau; ~~~\alpha = -\frac{{\dot x}_6}{ \Lambda} 
  =  \frac{{\dot x}^5}{ \Lambda} = \frac{1}{ \Lambda}.
\lbl{A3}
\ee
The action (\ref{A2}) can then be written as
\be
  I =\int \dd \tau \left [ p_\mu {\dot x}^\mu +p_5 
   - \frac{1}{2 \Lambda}(p_\mu p^\mu - M^2) \right ].
\lbl{A4}
\ee
Here $\Lambda$ is not a Lagrange multiplier, but a fixed quantity, namely $\Lambda \equiv - p_6$.

But we can omit $p_5$ in the above action, because it does not contribute to the $x^\mu$ equations
of motion. Then we have
\be
  I =\int \dd \tau \left [ p_\mu {\dot x}^\mu
  -\frac{1}{2 \Lambda}(p_\mu p^\mu - M^2) \right ].
\lbl{A5}
\ee
The corresponding Hamiltonian is
\be
  H = \frac{1}{2 \Lambda}(p_\mu p^\mu   - M^2)
\lbl{A6}
\ee
The above action is the Stueckelberg action. It is derived from the higher dimensional action.

From the constraint $p_M p^M - M^2 = p_\mu p^\mu    - 2 p_5 p_6 - M^2$
we have
\be
  p_5 = \frac{1}{2 p_6} (p_\mu p^\mu - M^2) = - H,
\lbl{A7}
\ee
which means that the Hamiltonian is given by the fifth component of momentum, and is
thus a generator of translations along $x^5 =\tau$, whilst the constant $\Lambda$ is given
by the sixth component of momentum.

Now let us do the same for a brane.
Let $\xi \equiv \xi^a$, $a=1,2,...,d$,  be $d$ parameters of a brane
in $(D+2)$-dimensions. Now a brane need not be space like. It can extend either into space like
or into time like directions, or both\,\ci{PavsicBook}. If $D=4$ then the extra two dimensions
are $x^5$ and $x^6$, but we may keep the same notation for the extra two dimensions even if
$D>4$. We then have $x^M = (x^\mu,x^5,x^6)$, $\mu = 0,1,2,3,7,8,...,D-3$.

The phase space brane action is
\bear
  &&I = \int \dd \tau  \dd^d \xi \left [ p_M {\dot x}^M 
  - \frac{\alpha}{2 \kappa \sqrt{{-\gam}}}(p_M p^M - \kappa^2 ({-\gam} )\right ]\nonumber\\
&&~~=  \int \dd \tau  \dd^d \xi \left [p_\mu {\dot x}^\mu +p_5 {\dot x}^5 +p_6 {\dot x}^6
- \frac{\alpha}{2 \kappa \sqrt{{-\gam}}}(p_\mu p^\mu    - 2 p_5 p_6 - \kappa^2 ({-\gam} )\right ].
\lbl{A8}
\ear
Choosing a gauge $\tau = x^5$, and using
\be
p_6 = - \sqrt{-\gam} {\tl \Lambda};~~~  {\dot x}_M = \frac{\alpha}{ \kappa \sqrt{{-\gam}}} p_M~;
{\dot x}_6 =  \frac{\alpha}{ \kappa \sqrt{{-\gam}}}(- \sqrt{-\gam} {\tl\Lambda}) = - {\dot x}^5  =-1,
\lbl{A10}
\ee
we obtain
\be
  I=   \int \dd \tau  \dd^d \xi \left [p_\mu {\dot x}^\mu -p_5  
-\frac{1}{2 {\sqrt{-\gam} \tl \Lambda}}(p_\mu p^\mu  - \kappa^2 ({-\gam})) \right ].
\lbl{A11}
\ee
Let us omit $p_5$, because this term does no influence the $x^\mu$ equations of motion.
Then we obtain the following unconstrained (Stueckelberg) action,
\be
  I=   \int \dd \tau  \dd^d \xi \left [p_\mu {\dot x}^\mu  
- \frac{1}{2 {\sqrt{-\gam} \tl \Lambda}}(p_\mu p^\mu  - \kappa^2 ({-\gam})) \right ]
\lbl{A12}
\ee 
which is a  generalized of the Stueckelberg point particle action.
The corresponding Hamiltonian is
\be
  H = \int \dd^d \xi (p_\mu {\dot x}^\mu - L )
 = \int \dd^d \xi \frac{1}{2 \sqrt{-\gam} {\tl \Lambda}}(p_\mu p^\mu  - \kappa^2 ({-\gam})) 
 = \int \dd^d \xi \, {\cal H} .
\lbl{A13}
\ee

From the constraint $p_\mu p^\mu    - 2 p_5 p_6 - \kappa^2 ({-\gam}) = 0$ we have
\be
  p_5 = \frac{1}{2 p_6} (p_\mu p^\mu  - \kappa^2 ({-\gam})) =  - {\cal H} .
\lbl{A14}
\ee

The Hamiltonian of a brane segment is
\be
   h = \Delta \xi \frac{1}{2 \sqrt{-\gam} {\tl \Lambda}}(p_\mu(\xi) p^\mu(\xi)  - \kappa^2 ({-\gam})) .
\lbl{A15}
\ee
Here  $p_\mu (\xi)$ is the momentum density. We introduce the momentum and the mass of a brane
segment
\be
p_\mu = p_\mu (\xi) \Delta \xi ~,~~~~~m = \kappa \sqrt{-\gam}\Delta \xi 
\lbl{A16}
\ee
We also define
\be
  \frac{p_\mu (\xi)}{\tl \Lambda} = \frac{p_\mu}{\Lambda} = \frac{p_\mu (\xi) \Delta \xi}{\Lambda}
\lbl{A17}
\ee
from which it follows
\be
  \Lambda = {\tl \Lambda } \Delta \xi
\lbl{A18}
\ee
 
The Hamiltonian of a brane segment thus becomes
\be
   h =  \frac{1}{2 {\Lambda}}(p_\mu p^\mu  - m^2)
\lbl{A19}
\ee

Equation of motion derived from the Stueckelberg brane action (\ref{A12}) is
\be
  \frac{\dd p_\mu (\xi)}{\dd \tau} 
  +\p^a \left [ \frac{1}{2 {\tl \Lambda}} \left ( \frac{1}{ \sqrt{-\gam}} p^\mu p_\mu 
  -\kappa^2 \sqrt{-\gam} 
 \right ) \p^a x_\mu \right ] = 0
\lbl{A20}
\ee
This corresponds to the quantum expectation value equation (\ref{5.44}).

\vs{4mm}

\end{document}